# CHALMERS

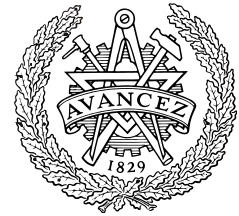

# Non-uniform sampling and reconstruction of multi-band signals and its application in wideband spectrum sensing of cognitive radio

MOSLEM RASHIDI

*Signal Processing Group*
*Department of Signals and Systems*
Chalmers University of Technology

Göteborg, Sweden, 2010                                                        EX038/2010

# Abstract

Sampling theories lie at the heart of signal processing devices and communication systems [1]. To accommodate high operating rates while retaining low computational cost, efficient analog-to digital (ADC) converters must be developed [1]. Many of limitations encountered in current converters are due to a traditional assumption that the sampling state needs to acquire the data at the Nyquist rate, corresponding to twice the signal bandwidth [1].

In this thesis a method of sampling far below the Nyquist rate for sparse spectrum multiband signals is investigated. The method is called periodic non-uniform sampling, and it is useful in a variety of applications such as data converters, sensor array imaging and image compression.

Firstly, a model for the sampling system in the frequency domain is prepared. It relates the Fourier transform of observed compressed samples with the unknown spectrum of the signal. Next, the reconstruction process based on the topic of compressed sensing is provided. We show that the sampling parameters play an important role on the average sample ratio and the quality of the reconstructed signal. The concept of condition number and its effect on the reconstructed signal in the presence of noise is introduced, and a feasible approach for choosing a sample pattern with a low condition number is given. We distinguish between the cases of known spectrum and unknown spectrum signals respectively.

One of the model parameters is determined by the signal band locations that in case of unknown spectrum signals should be estimated from sampled data. Therefore, we applied both subspace methods and non-linear least square methods for estimation of this parameter. We also used the information theoretic criteria (Akaike and MDL) and the exponential fitting test techniques for model order selection in this case.

In the area of spectrum sensing for cognitive radio, there is a tendency towards the wideband sensing. The main bottleneck for this desire is the requirement of a high sample rate ADC. Hence, we propose a model for the wideband spectrum sensing from non-uniform samples that are taken by a low rate non-uniform ADC. Depend on the application, the wideband of interest is divided into a finite number of channels and the presence of a primary user in each channel is examined. We show how to design and specify the model parameters. Also we evaluate our model performance by computing the detection probability in terms of the SNR and compression ratio.

# Acknowledgment

I wish to thank Professor Mats Viberg and Professor Lars Svensson for providing this opportunity to work under their supervision. I give special thanks to them for introducing this field of research to me and for their helpful comments and insights throughout the thesis that makes me deeper, more efficient and productive. I also wish to thank Arash Owrang and Ashkan Panahi for the good discussions and great times we had during the work. Finally, I wish to thank Kasra Haghighi for his help in the spectrum sensing part for cognitive radio.

# Contents

# 1. Background and Theory

## 1.1 Introduction

Reception and reconstruction of analog signals are performed in a wide variety of applications, including wireless communication systems, spectrum management applications, radar systems, medical imaging systems and many others. In many of these applications, an information-carrying analog signal is sampled, i.e., converted into digital samples. The information is then reconstructed by processing the digital samples [35].

The classical sampling theorem states that a real low-pass signal, band limited to the range ($-f_{max}$, $+f_{max}$) can be reconstructed from its uniform samples, provided the sampling rate satisfies the Nyquist rate that is $f_{nyq}=2f_{max}$.

While the uniform sampling theorem is suitable for low-pass signals and an efficient sampling with minimum rate is attained, it seems quite inefficient in case of signals with multiple bands with sparse spectrum [3]. These signals do not occupy the whole frequency band and uniform sampling can become very redundant [4]. This comes from the fact that multiband signals have some gaps between each band that tempts one to work with a rate lower than Nyquist rate [3]. Following this vision, a clever way of sampling the signal that is called "multicoset sampling" or "periodic non-uniform sampling" is used at a rate lower than the Nyquist rate, that captures enough information to enable perfect reconstruction.

This project studies the periodic non-uniform sampling and reconstruction of multiband sparse spectrum signals. The outline of the article is as follows: first the signal model and some of the definitions are described. Then the Non-uniform sampling method is introduced and the reconstruction model is expressed. In Section II, the discussion focuses on the known spectrum signals to find the suitable parameters and their effect on the reconstruction of the signal. Section III covers the spectral recovery of unknown spectrum signals. In Section IV we used the idea of non-uniform sampling in the spectrum sensing of cognitive radio systems. The last part is the summary and conclusions.

## 1.2 Signal Model and Definitions

Let $B(F)$ denote the class of continuous complex-valued signals of finite energy, bandlimited to a subset $F$ of the real line (consisting of a finite union of bounded intervals) [5][3]:

$$B(F) = \{x(t) \in L^2(\mathbb{R}) \cap C(\mathbb{R}): X(f) = 0, f \notin F\}$$

(1.1)

where

$$F = \bigcup_{i=1}^{N}[a_i, b_i)$$

(1.2)

the subset $F \subset [0, f_{max}]$ is called the spectral support of signal, and

$$X(f) = \int_{-\infty}^{\infty} x(t)\exp(-j2\pi ft)dt$$

(1.3)

is the *Fourier* transform of $x(t)$.

To quantify the sampling efficiency for signals with a given spectral support $F$, we define the spectral occupancy as

$$\Omega = \frac{\lambda(F)}{f_{max}}, \quad 0 \leq \Omega \leq 1$$

(1.4)

where $f_{max}$ is the highest frequency and $\lambda(.)$ denotes the *Lebesgue* measure [3]. The Lebesgue measure is the standard way of assigning a length, area or volume to subsets of Euclidean space [6]. The Lebesgue measure for the set $F$ defined in (1.2) is

$$\lambda(F) = \sum_{i=1}^{N}(b_i - a_i)$$

(1.5)

The *Nyquist* rate for signals with spectral support $F$ is defined as the smallest uniform sampling rate that guarantees no aliasing [3]

$$f_{nyq} = \inf\{\theta > 0 : F \cap (n\theta \oplus F) = \emptyset, \forall n \in Z \setminus \{0\}\}$$

(1.6)

where

$$\theta \oplus F = \{\theta + f : f \in F\}$$

(1.7)

is the translation of the set $F$ by $\theta$. Then, the *Nyquist* sampling rate satisfies

$$\lambda(F) \leq f_{nyq} \leq f_{max}$$

(1.8)

We say that $F$ is packable if $f_{nyq} = \lambda(F)$, and nonpackable otherwise ($f_{nyq} > \lambda(F)$). The general case of interest is when the signal is totally nonpackable, that is $f_{nyq} = f_{max}$ [3].

Landau [7] showed that the sampling rate of an arbitrary sampling scheme for the class of multiband signals with spectral support $F$ is lower-bounded by the quantity $\lambda(F)$, which may be significantly smaller than the *Nyquist* rate. Thus the spectral occupancy is a measure of the efficiency of Landau's lower bound over the *Nyquist* rate. Because $\Omega$ can be low for certain nonpackable signals, uniform sampling is highly inefficient for such signals.

Fig. 1.1 illustrates a typical case of such a nonpackable multiband signal. The spectral support is $F=\{[0.5,2],[4,5],[8,8.5]\}$, the *Nyquist* rate for this signal is $f_{nyq}=f_{max}=12$ (hence it is totally nonpackable), whereas the Landau lower bound from (1.5) is

$\lambda(F)=(2-0.5)+(5-4)+(8.5-8)=1.5+1+.5=3$ .

The spectral occupancy from (1.4) for this signal, Ω=3/12=0.25, suggests that it might be possible to sample the signal four times as efficiently as the Nyquist rate [3].

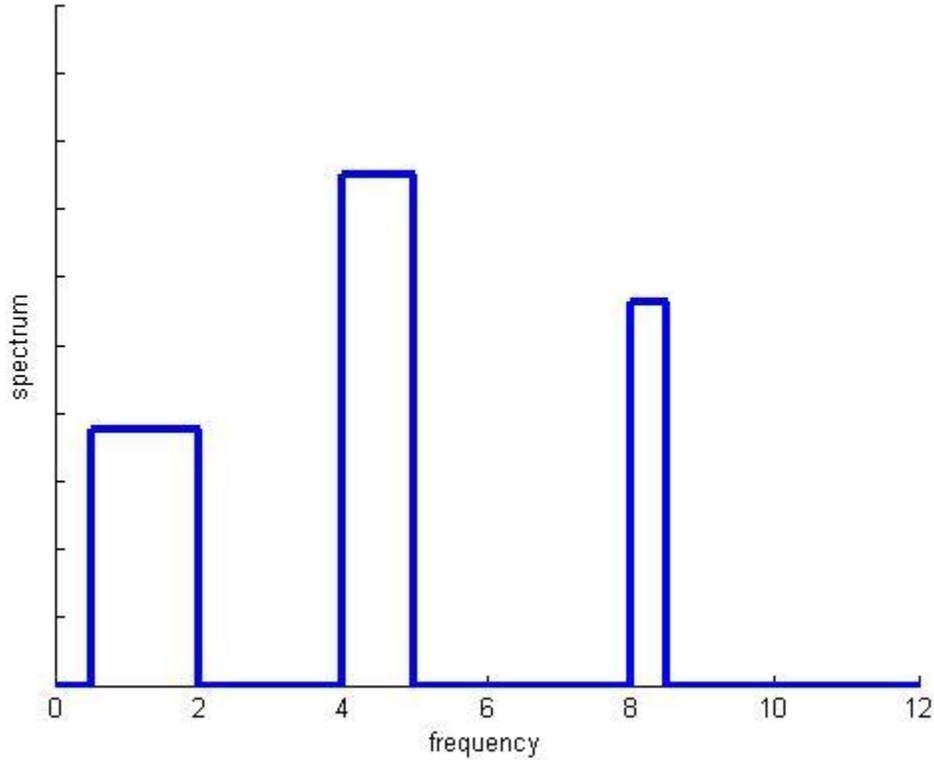

Fig.1.1: Spectrum of a multiband signal, with $f_{max}$=12, N=3, F={[0.5,2],[4,5],[8,8.5]}

## 1.3     Non-uniform sampling

Uniform sampling is not well suited for nonpackable signals. However, it turns out that there is a clever way of sampling the signal *x(t)* called "multi-coset sampling" or "periodic non-uniform sampling" at a rate lower than the *Nyquist* rate, that captures enough information to recover *x(t)* exactly [3].

Let *x(t)* ∈ *B(F)*. In multi-coset sampling, we first pick a suitable sampling period $T$ (such that uniform sampling at rate *1/T* causes no aliasing), and a suitable integer *L > 0*, and then sample the input signal *x(t)* non-uniformly at the instants $t_i(n) = (nL + c_i)T$ for $1 \leq i \leq p$ and $n \in \mathbb{Z}$, (Fig.1.2). The set $\{c_i\}$ contains *p* distinct integers chosen from set $\mathbb{L} = \{0,1, \dots, L-1\}$.

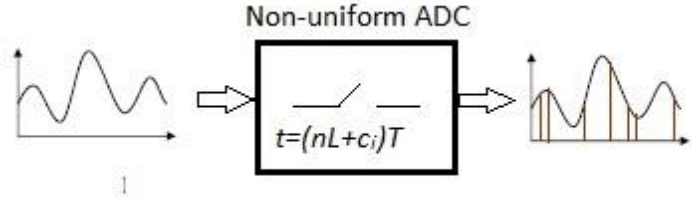

Fig.1.2: Multi-coset sampling operation

This process of sampling can be viewed as first sampling the signal at the "base sampling rate" of *1/T*, and then discarding all but *p* samples in every block of *L* samples periodically. The samples that are retained in each block are specified by the set $\{c_i\}$ [3].

For a given $c_i$, the coset of sampling instants $t_i(n) = (nL + c_i)T$, $n \in \mathbb{Z}$ is uniform with inter-sample spacing equal to *LT*. We call this the *i-th* active coset. The set $C=\{c_i\}$ is referred to as the (*L,p*) sampling pattern and the integer *L* as the period of pattern where [3],[8]

$$0 \leq c_1 \leq c_2 \leq \ldots \leq c_p \leq L - 1.$$

(1.9)

Fig.1.3 shows two mulicoset sampling patterns corresponding to parameters (*L,p*)=(20,5). In the Fig 1.3(a), when *n=0* the first block of samples at times $t_i(0)=\{0,4,7,12,16\}$, and when *n=1* the the second block of samples of times $t_i(1)=\{20,24,27,32,36\}$ are kept. The corresponding sampling times for Fig 1.3(b) are $t_i(0)=\{2,6,11,15,18\}$ and $t_i(1)=\{22,26,31,35,38\}$ respectively.

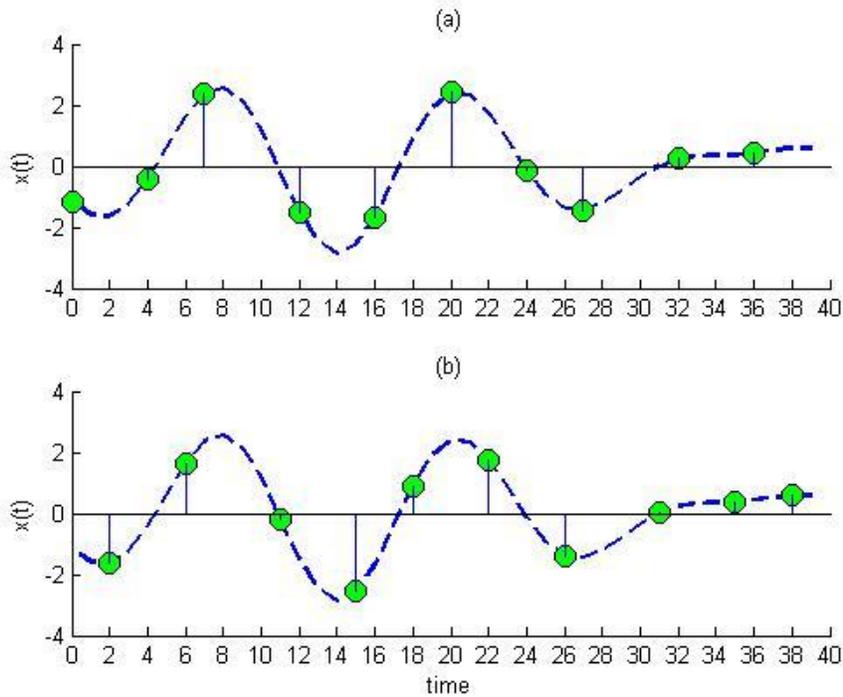

Fig.1.3: Two different sampling patterns for *(L,p)=(20,5)*, (a) *C={0,4,7,12,16}*.(b) *C={2,6,11,15,18}*.

One possible implementation of the Non-uniform ADC is illustrated in Fig.1.4. It is composed of $p$ parallel ADCs, each working uniformly with a period of $T_s = LT$ and a sampling time offset by $\{c_i T\}$. The clock generator block takes the input sample clock $T_s$ and provides the required $p$ sample clocks for each ADC according to the sampling pattern such that

$$t_i(n) = (nL + c_i)T = nLT + c_i T = nT_s + \frac{c_i}{L}T_s = \left(n + \frac{c_i}{L}\right)T_s, \quad 1 \leq i \leq p$$

(1.10)

The implementation follows the structure of interleaved ADC converters with the exception that in the interleaving converters, the ADCs are triggered sequentially.

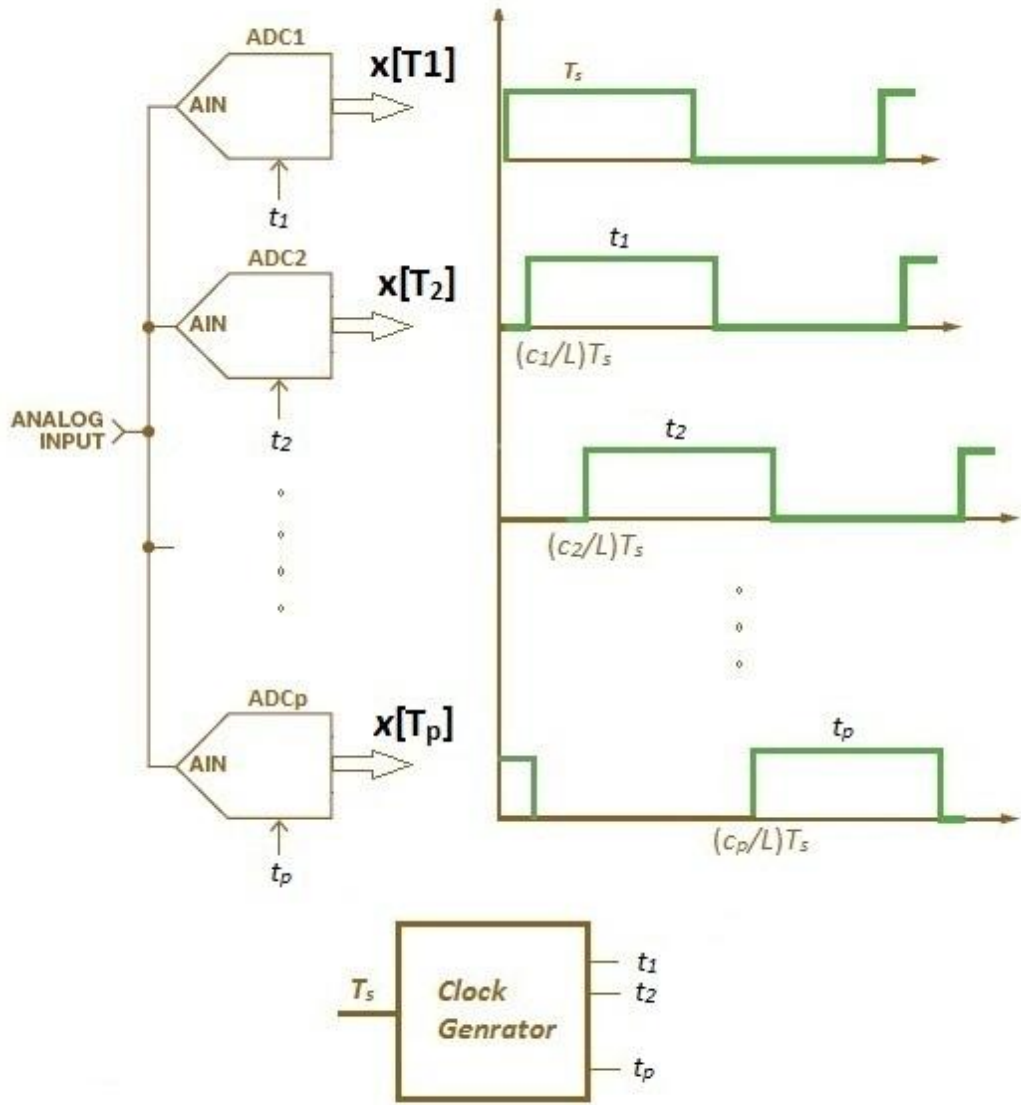

Fig.1.4: A parallel implementation of Non-uniform ADC and its clock timing

Define the *i-th* sampling sequence for $1 \leq i \leq p$ as [8]
$$x_i[n] = \begin{cases} x[nT], & n = mL + c_i, m \in \mathbb{Z} \\ 0, & otherwise \end{cases}$$
(1.11)

The sequence of $x_i[n]$ is obtained by up-sampling the output of the *i-th* ADC with a factor of $L$ and shifting in time with $c_i$ samples. Fig.1.5 shows the sequences of $x_1[n]$, $x_2[n]$, $x_4[n]$, for the signal of Fig 1.2(b).

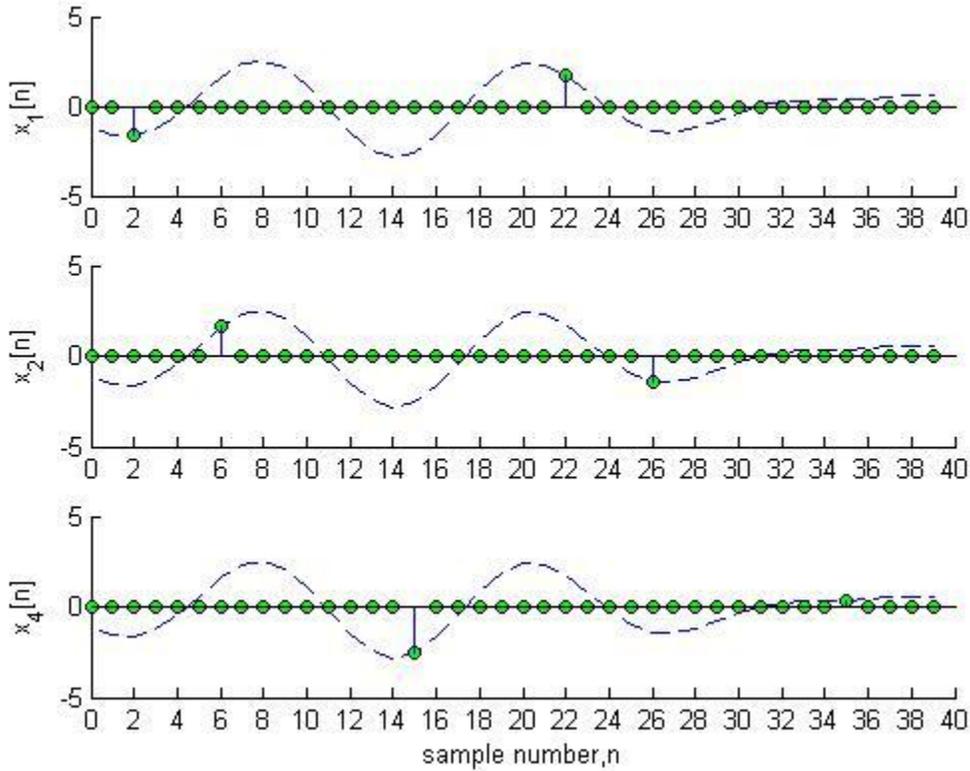

Fig.1.5: The first, second and forth sampling sequences $x_1[n]$,$x_2[n]$,$x_4[n]$, with *p=5*

Direct calculations link the known discrete-time Fourier transform $X_i(e^{j2\pi fT})$ of $x_i[n]$ to the unknown Fourier transform $X(f)$ of $x(t)$ [8]:

$$X_i(e^{j2\pi fT}) = \sum_{n=-\infty}^{+\infty} x_i[n] \exp(-j2\pi fnT)$$

$$= \frac{1}{LT} \sum_{r \in \mathbb{Z}} \exp\left(j\frac{2\pi}{L} c_i r\right) X\left(f + \frac{r}{LT}\right)$$
(1.12)

which, using the fact that X(*f*)=0 for $f \notin [0, f_{max}]$, gives us [3]

$$X_i(e^{j2\pi fT}) = \frac{1}{LT}\sum_{r=0}^{L-1} \exp\left(j\frac{2\pi}{L}c_i r\right) X\left(f + \frac{r}{LT}\right)$$

(1.13)

for every $1 \leq i \leq p$, and every $f$ in the interval [8]

$$F_0 = [0, \frac{1}{LT})$$

(1.14)

Let us express (1.13) in matrix form as

$$\mathbf{y}(f) = \mathbf{A}_C\, \mathbf{s}(f), \quad \forall f \in F_0$$

(1.15)

where $\mathbf{y}(f)$ is a vector of length $p$ whose $i$-th element is $X_i(e^{j2\pi fT})$,

$$\mathbf{y}(f)_{p\times 1} = \begin{bmatrix} X_1(e^{j2\pi fT}) \\ X_2(e^{j2\pi fT}) \\ \vdots \\ X_p(e^{j2\pi fT}) \end{bmatrix}, \forall f \in F_0$$

and $\mathbf{A}_C$ is a $(p \times L)$ matrix with $il$-th element given by [5],[8]

$$\mathbf{A}_C(i, l+1) = \frac{1}{LT}\exp\left(\frac{j2\pi c_i l}{L}\right), \quad 1 \leq i \leq p,\, 0 \leq l \leq L-1$$

(1.16)

Note that $\mathbf{A}_C$ is a sub-matrix of the complex conjugate of the $L \times L$ discrete Fourier Transform (DFT) matrix, consisting of the $p$ rows indexed by the sampling pattern $C$ [5]. According to (1.16) $\mathbf{A}_C$ is given by

$$\mathbf{A}_{C\,p\times L} = \frac{1}{LT}\begin{bmatrix} 1 & e^{\frac{j2\pi c_1}{L}} & \ldots & \ldots & e^{\frac{j2\pi(L-1)c_1}{L}} \\ 1 & e^{\frac{j2\pi c_2}{L}} & \ldots & \ldots & e^{\frac{j2\pi(L-1)c_2}{L}} \\ \vdots & \vdots & \ldots & \ldots & \vdots \\ \vdots & \vdots & \ldots & \ldots & \vdots \\ 1 & e^{\frac{j2\pi c_p}{L}} & \ldots & \ldots & e^{\frac{j2\pi(L-1)c_p}{L}} \end{bmatrix}$$

(1.17)

and the vector $\mathbf{s}(f)$ contains $L$ unknowns as [5],[8]

$$\mathbf{s}(f)_{L\times 1} = \begin{bmatrix} X(f) \\ X(f+\frac{1}{LT}) \\ \vdots \\ X(f+\frac{L-1}{LT}) \end{bmatrix}, \forall f \in F_0$$

(1.18)

The relation (1.18) states that the unknown elements of vector $s(f)$ are created by first bandpass filtering of the original signal to the range $\frac{r}{LT} \leq f < \frac{r+1}{LT}$, and then frequency shifting to the left by $\frac{r}{LT}$ units [3]. In the other words, if the spectrum of signal, X($f$) is sliced into $L$ cells indexed from 0 to $L-1$, then each cell corresponds to the associated row of the vector $s(f), \forall f \in F_0$. Fig.1.6 illustrates the spectrum of a typical multiband signal that is sliced into $L=5$ cells indexed from 0 to 4, the third element of $s(f)$ that is indexed by 2 is highlighted in the figure. Denoting the inverse Fourier transform of $X\left(f + \frac{r}{LT}\right)$ by $x_r(t)$, it is evident from the above definition that

$$x(t) = \sum_{r=0}^{L-1} x_r(t) \exp\left(\frac{j2\pi rt}{LT}\right)$$

(1.19)

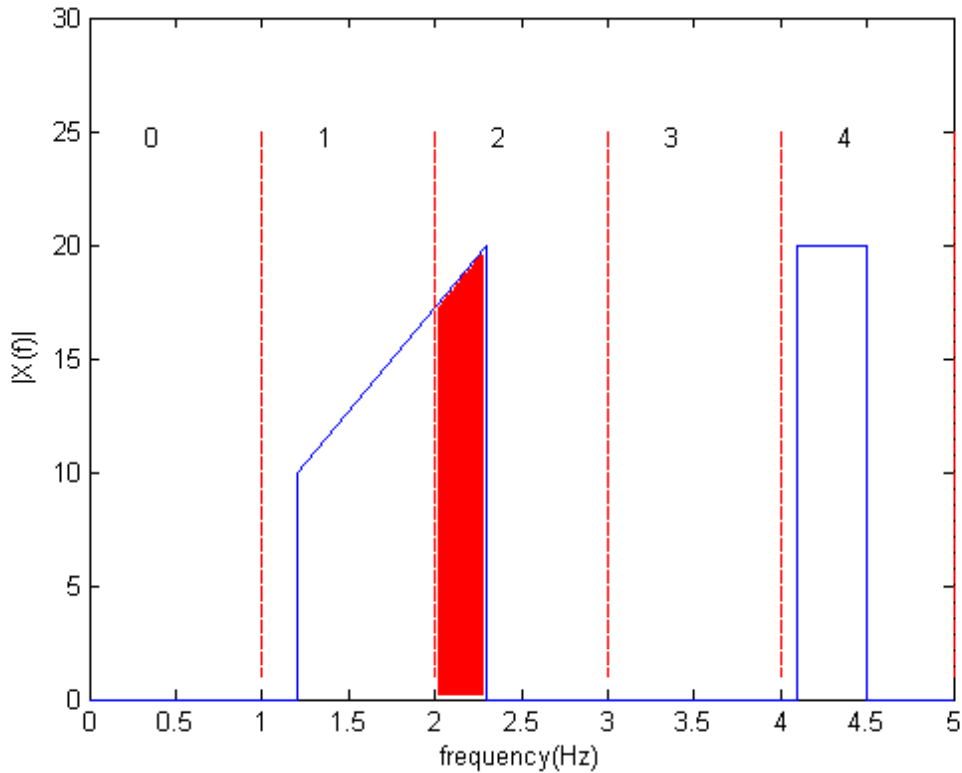

Fig.1.6: Frequency representation of a 2-bands signal that is sliced into $L=5$ cells. The spectral index set is $k=\{1,2,4\}$

The problem of recovering $x(t)$ is then equivalent to solving the linear system of equations (1.15) for every $f \in F_0$. This process is explained in the next section.

## 1.4 Reconstruction

The goal of the reconstruction scheme is to perfectly recover $x(t)$ from the set of sequences $x_i[n]$, $1 \leq i \leq p$, or equivalently, to reconstruct $\mathbf{s}(f)$ for every $f \in F_0$ from the input data $\mathbf{y}(f)$ [8]. The relation (1.15) states that for each $f \in F_0$, the vector $\mathbf{y}(f)$ has $p$ known elements while the vector $\mathbf{s}(f)$ has $L$ unknown elements and as $p < L$, then the number of equations is less than the number of unknowns [5][8]. This is the case in the compressed sensing problem that we briefly review it below.

Compressive sampling (CS) is a method for acquisition of sparse signals at rates significantly lower than Nyquist rate. Let the analog sparse signal $x(t)$, be represented as a finite weighted sum of basis functions (e.g. Fourier) $\psi_i(t)$ as follows

$$x(t) = \sum_{i=1}^{N} s_i \psi_i(t)$$

(1.20)

where only a few basis coefficients $s_i$ are non-zero due to sparsity of $x(t)$ [27].
In a discrete time framework, $N$ samples of $x(t)$ in a $\mathbf{N \times 1}$ vector can be represented in matrix form as

$$\mathbf{x} = \mathbf{\psi s}$$

(1.21)

where $\mathbf{\psi}$ is the $\mathbf{N \times N}$ representation basis matrix with $\psi_1,\ldots, \psi_N$ as columns and $\mathbf{s}$ is an $\mathbf{N \times 1}$ vector with $K \ll N$ non-zero entries $s_i$ [28].

The samples of $x(t)$ in a standard form are given by

$$y_k = <x, \boldsymbol{\phi}_k>, \quad k \in M, M \subset \{1, \ldots, N\},$$

(1.22)

where $\boldsymbol{\phi}_k(t)$ is the sensing waveform. If the sensing waveforms are Dirac delta functions (spikes), for example, then $\mathbf{y}$ is a vector of sampled values of $x$ in the time or space domain [28]. The measurement vector $\mathbf{y}$ can be written in matrix form as

$$\mathbf{y} = \boldsymbol{\phi} \mathbf{x} = \boldsymbol{\phi} \boldsymbol{\psi} \mathbf{s}$$

(1.23)

where $\boldsymbol{\Phi}$ is the $\mathbf{M \times N}$ sensing basis matrix that is, in general, incoherent with $\boldsymbol{\psi}$. An example construction of $\boldsymbol{\Phi}$ is by choosing elements that are drawn independently from a random distribution, e.g., Gaussian or Bernoulli.

The reconstruction is achieved by solving the following $l_1$-norm optimization problem

$$\hat{\mathbf{s}} = \arg\min_{\mathbf{s}} \|\mathbf{s}\|_1 \quad s.t. \quad \mathbf{y} = \boldsymbol{\phi} \boldsymbol{\psi} \mathbf{s}$$

(1.24)

It follows that with the proposed multi-coset periodic sampling scheme, the sampling problem for continuous-time signals has been reduced, for each fixed $f$, to a CS problem for signals in $\mathbb{C}^L$ that are sparse in the DFT sense [5].

Therefore by using the fact that some cells are free of energy, the number of unknowns can be reduced in (1.15). Fig.1.6 shows the spectrum of a multi-band signal where the cells indexed by 1, 2 and 4 contain nonzero parts of the spectrum and the cells with indices 0 and 3 are free from energy. Hence, it is enough to solve (1.15) for only three unknowns rather than five. The cells of signal that contains the nonzero part of spectrum are termed active cells. Denote the number of such active cells by $q \leq L$ [5]. To reduce the order of equation (1.15) we need to know which cells those are active. The set $\boldsymbol{k}$ is defined as the spectral index set of the signal [8], and it indicates the cells that are nonzero, such that

$$\boldsymbol{k} = [k_1, k_2, \ldots, k_q], k_r \in \mathbb{L}, 1 \leq r \leq q,$$

(1.25)

Define the reduced signal vector

$$\boldsymbol{z}(f) = \begin{bmatrix} X(f + \frac{k_1}{L}) \\ X(f + \frac{k_2}{L}) \\ \vdots \\ X(f + \frac{k_q}{L}) \end{bmatrix} \in \mathbb{C}^q$$

(1.26)

that contains only the $q$ active cells indexed by the set $\boldsymbol{k}$, and the reduced measurement matrix $\mathbf{A}_C(\boldsymbol{k}) \in \mathbb{C}^{p \times q}$ is derived by choosing the columns of $\mathbf{A}_C$ that are indexed by the spectral index $\boldsymbol{k}=\{k_1, k_2, \ldots, k_q\}$.

$$\mathbf{A}_c(\boldsymbol{k})_{p \times q} = \frac{1}{LT} \begin{bmatrix} e^{\frac{j2\pi k_1 c_1}{L}} & e^{\frac{j2\pi k_2 c_1}{L}} & \ldots & \ldots & e^{\frac{j2\pi k_q c_1}{L}} \\ e^{\frac{j2\pi k_1 c_2}{L}} & e^{\frac{j2\pi \times k_2 c_2}{L}} & \ldots & \ldots & e^{\frac{j2\pi k_q c_2}{L}} \\ \vdots & \vdots & \ldots & \ldots & \vdots \\ \vdots & \vdots & \ldots & \ldots & \vdots \\ e^{\frac{j2\pi k_1 c_p}{L}} & e^{\frac{j2\pi k_2 c_p}{L}} & \ldots & \ldots & e^{\frac{j2\pi k_q c_p}{L}} \end{bmatrix}$$

(1.27)

Equation (1.15) then reduces to [5],[9]

$$\boldsymbol{y}(f) = \mathbf{A}_C(\boldsymbol{k})\boldsymbol{z}(f), \quad \forall f \in F_0$$

(1.28)

If $\mathbf{A}_C(\boldsymbol{k})$ has full column rank, the unique solution can be obtained using a left inverse, e.g. the pseudo-inverse of $\mathbf{A}_C(\boldsymbol{k})$ that we simply denote by $\mathbf{A}_C^\dagger$ [5],[9]:

$$\boldsymbol{z}(f) = \boldsymbol{A}_C^\dagger \, \boldsymbol{y}(f) \, , \quad \forall f \in F_0$$

(1.29)

After finding **z**(*f*), the time domain representation of each cell is achieved by taking inverse Fourier transform, and with proper combination according to (1.19) the signal, *x(t)* is reconstructed.

A simple time domain solution for the recovery of *x(t)* involves filtering of the sequences $x_i[n], i=1,...p$ to produce $x_{hi}[n]$ and linear combination of filtered sequences using $A_C^\dagger$ producing *x(nT)* [5],[9]. The interpolation filter *h[n]* with cut off frequency at $f_c = \frac{f_{max}}{L}$ filters the sequence $x_i[n]$ that is upsampled with *L* as defined in (1.11), i.e.

$$x_{hi}[n] = h[n] * x_i[n]$$

(1.30)

The reconstruction formula is then

$$x(nT) = \sum_{i=1}^{q}\sum_{l=1}^{p}[A_C^\dagger]_{il}\, x_{hl}[n] \exp\left(j\frac{2\pi k_i n}{L}\right)$$

(1.31)

This is the Nyquist-rate sampled version of the desired continuous-time signal *x(t)*, so that *x(t)* can be recovered by a standard D/A [9]. Fig.1.7 shows the reconstruction of *x(t)* from sequences $x_i[n]$. All filters have the same low pass response, which is advantageous for implementation. The coefficients are $a_i = \exp\left(\frac{j2\pi n k_i}{L}\right)$

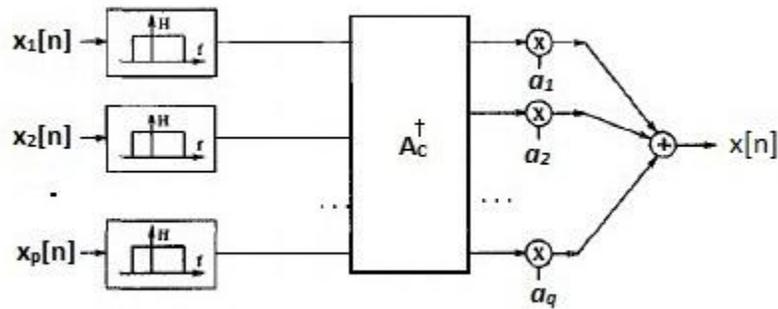

Fig.1.7: Reconstruction of uniformly sampled x[n] from the non-uniformly sampled $x_i[n]$ sequences

## 2. Known Spectral Support Signals

If the band locations of the multiband signal are given as (1.2) we encounter with the known spectrum case. The spectral index set and the sampling parameters can be obtained by exploiting this prior information about the spectrum of signal. The process is described as follows:

### 2.1 Spectral index set $k$

As previously mentioned the spectrum of the signal is divided into $L$ cells with width of $\frac{f_{max}}{L}$ and indexed from 0 to $L-1$. Therefore, each band of the signal located at $[a_i, b_i)$ as Fig. 2.1 shows can be overlapped by grouping of cells indexed by

$$\left\lfloor \frac{a_i * L}{f_{max}} \right\rfloor \leq k_i \leq \left\lfloor \frac{b_i * L}{f_{max}} \right\rfloor, \quad 1 \leq i \leq N$$

(2.1)

where $\lfloor \ \rfloor$ is floor function and $k_i$ is the set of indices for each band. After finding the set $k_i$ for all bands, the spectral index set is

$$k = \bigcup_{i=1}^{N} k_i$$

(2.2)

The number of active cells is the cardinality or length of the spectral index set

$$q = |k|$$
(2.3)

**Example**: The signal in Fig.1.6 has 2 bands, with spectral support of $F=\{[1.2,2.2),[4.1,4.5)\}$, with $L=5$, $f_{max}=5$, the cells that are occupied by each band are respectively

$\lfloor 1.2 * 5/5 \rfloor \leq k_1 \leq \lfloor 2.2 * 5/5 \rfloor \Rightarrow 1 \leq k_1 \leq 2 \Rightarrow k_1=\{1,2\}$
$\lfloor 4.1 * 5/5 \rfloor \leq k_2 \leq \lfloor 4.5 * 5/5 \rfloor \Rightarrow 4 \leq k_2 \leq 4 \Rightarrow k_2=\{4\}$
$k = \bigcup_{i=1}^{2} k_i = \{1,2,4\}$ and $q=3$

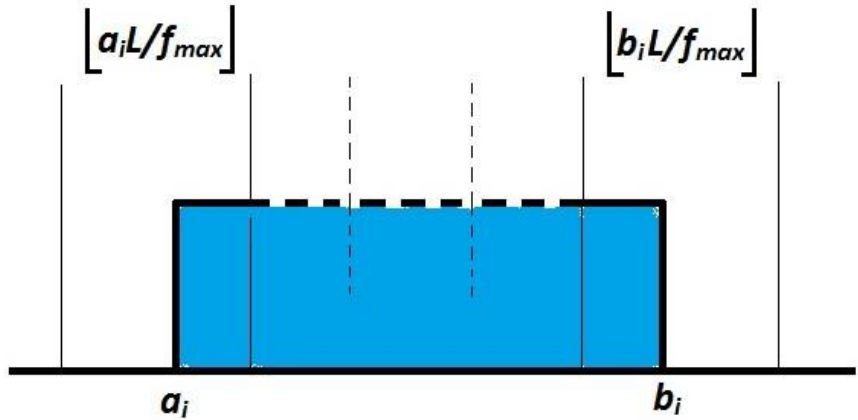

Fig: 2.1: The *i-th* band of the signal located at $[a_i,b_i]$ occupies a group of cells determined by (2.1)

## 2.2 Sampling Parameters

In the non-uniform sampling, the parameters *T, L, p* and *C* should be selected properly for a perfect and optimal reconstruction. The most useful criteria to choose these parameters are minimum sampling rate, minimum error and perfect reconstruction. In fact, it turns out that unless the sampling and reconstruction system is very carefully designed and optimized, the sensitivity to small errors can be so great that although perfect reconstruction is possible with perfect data, the signal is corrupted beyond recognition in most practical situations [11].We consider selecting these parameters in the following sections.

### 2.2.1 Base Sampling time *T*:

The base sampling rate *T* could be chosen equal to the Nyquist rate, i.e., $T=1/f_{nyq}$, but never lower. However, we choose

$$T = \frac{1}{f_{max}}$$

(2.4)

because sampling at this rate always guarantees no aliasing for any *F* [11].

### 2.2.2 *L* and *p*:

As we saw in Fig.1.4, the output of each ADC is periodic with the period of $T_s=LT$. This makes a lower bound for choosing *L* based on the capability of hardware and ADC sampling time such that

$$L > \frac{T_{ADC}}{T}$$

(2.5)

In a Non-uniform sampling with (*L,p*) parameters, the average sampling rate when choosing $T=1/f_{max}$ is [5]

$$\bar{f} = \left(\frac{p}{L}\right) f_{max}$$

(2.6)

Then, it is clear that a large *L* and small *p* is desired for a minimum sample rate close to Landau lower bound. The parameter *p* is selected with respect to the number of active cells *q*, such that $p \geq q$ to have enough known equations in (1.28). Therefore, choosing a large *L* introduces more active slots, and then it needs using higher *p* or equivalently higher number of ADCs and hardware according to Fig.1.4. Also, the computations for reconstruction depends directly on the dimensions of **y**(*f*) and $\mathbf{A}_C$ that grow with large *L* and *p*.

While it seems that a large *L* results in a lower sample rate, this is not a general rule. As (2.1) and (2.2) show, the number of active cells depends on *F*, *L* and *T*. Hence, we may choose a larger *L* and still have the same or even higher sample rate. Comparing Fig.2.2 (a) and (b): in the first case *L=5* and number of active cells is *q=2*, then $\bar{f} \geq \frac{2}{5}$. By increasing to *L=10*, the number of active cells is *q=4*, then $\bar{f} \geq \frac{4}{10} = \frac{2}{5}$. Thus, in this case increasing *L* and then *p* costs more in term of hardware and computation, but it gains no sampling rate reduction.

Therefore the parameters *L* and *p* in the known spectrum case could be optimized based on an intuitive consideration of computations, hardware capabilities and achieving minimum sampling rate with minimum value of *L* and *p*. Because small values of *L* may often suffice, and larger *L* increases the computation cost of the reconstruction of the signal, small to moderate values of *L* (e.g., in the tens to hundreds) are of interest [8]. When *L* is selected, *q* is obtained from (2.1)-(2.3) hence $p \geq q$ is selected. This condition also satisfies the Landau lower bound, that is $\bar{f} > \lambda(F)$.

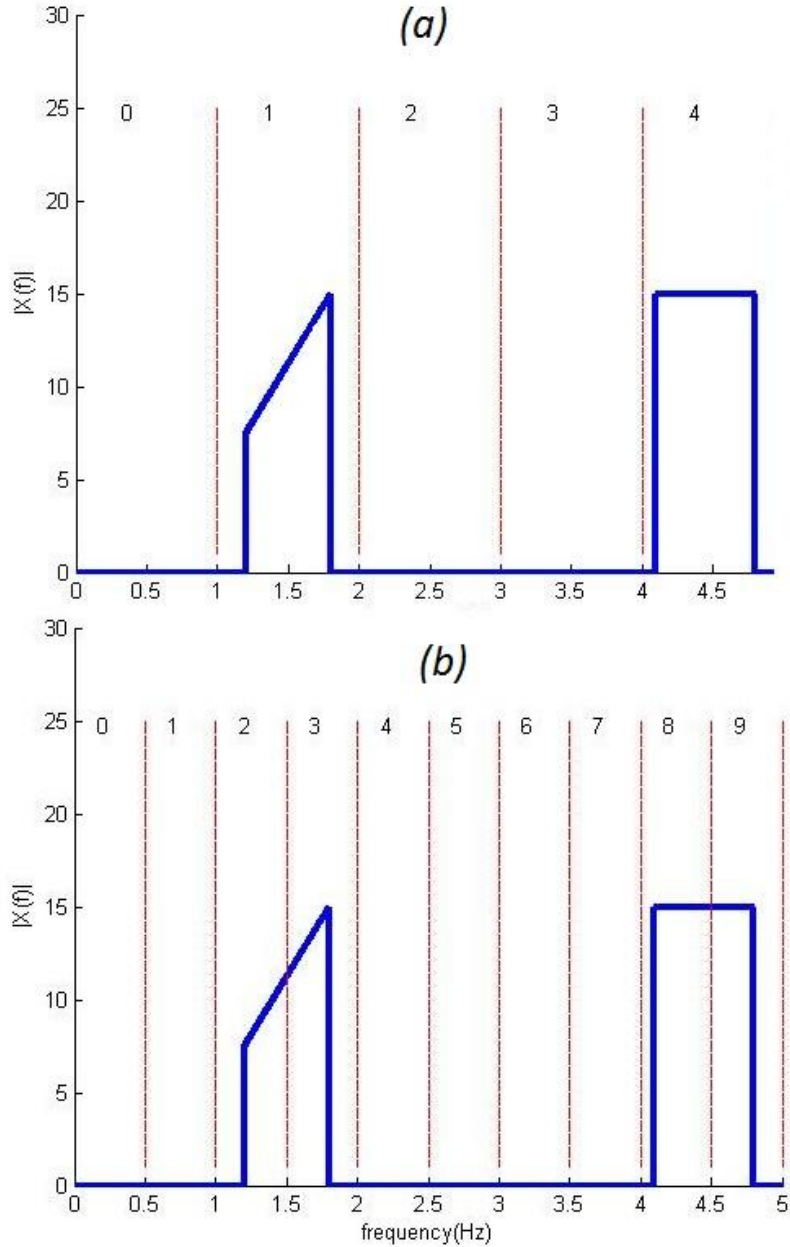

Fig. 2.2: Active cells and minimum sample rate (a) *L=5*, $\bar{f} = \frac{2}{5}$  (b) *L=10*, $\bar{f} = \frac{2}{5}$

### 2.2.3 Sample pattern *C*

The sample pattern is the selection of *p* out of *L* numbers from 0 to *L-1* that should be chosen after fixing *L* and *p*. We will see that finding a good sample pattern can optimize the aliasing error bounds and sensitivity to noise in the reconstruction process.

In the reconstruction part we saw that the pseudo-inverse of **A**$_C$(**k**) exists if **A**$_C$(**k**) is full column rank. Hence, a sample pattern *C* that yields a full column rank **A**$_C$, also called universal [8], results in **A**$_C$(**k**) full rank too. This is the first criterion for choosing the sample pattern *C*.

Since in practice the left hand side of equation (1.28) will be perturbed owing to *x(t)* being imperfectly band-limited to *F*, and due to quantization errors or phase noise, the numerical stability or conditioning of **A**$_C$(**k**) is a very important issue. Therefore, a sampling pattern that results in a well-conditioned **A**$_C$(**k**) is highly desirable as the second criterion [9].

A system of equations is considered to be well-conditioned if a small change in the coefficient matrix or a small change in the left hand side results in a small change in the solution vector [10]. This is achieved by choosing a coefficient matrix with low condition number. The condition number of matrix **A** is defined as

$$cond(\mathbf{A}) = \|\mathbf{A}\| \cdot \|\mathbf{A}^{-1}\|$$
$$= \frac{\sigma_{max}(\mathbf{A})}{\sigma_{min}(\mathbf{A})}$$

(2.7)

where ‖ ‖ is the norm operation and $\sigma_{max}$ and $\sigma_{min}$ are the maximum and minimum singular values respectively. An ideal sample pattern is defined to give the *cond*(A$_C$(**k**))=*1* among all patterns that are universal (at a fixed resolution *L*) for the target set of spectral supports [9]. However, depending on the spectral index set, **k** it is not possible to achieve condition number of one, then a pattern with the smallest *cond*(**A**$_C$(**k**)) is desired. Such a sampling pattern can be found as the solution to the following minimization problem [9]:

$$\mathbf{C}_{opt} = \arg \min_{C:|C|=p} cond\left(\mathbf{A_C}(\mathbf{k})\right)$$

(2.8)

where the symbol |*C*| gives the cardinality or length of the set *C*. Solution of (2.8) by exhaustive search would require $\binom{L}{p}$ evaluations of the condition of **A**$_C$(**k**), which is feasible only for small values of *L* and *p*. Invariance to circular shifts and mirroring (modulo *L*) of *C* and **k** can be used to reduce the search [5].

For a typical case when *L=16, p=5* and **k**={3,4,5,10,11} with $\binom{16}{5} = 4368$ evaluations, the result in table I is achieved. The worst pattern has a huge condition that can explode the result. Also, a random pattern and a bunch pattern which contains *p* consecutive numbers such as *C={0,1,...,p-1}* are created to compare with maximum and minimum values. The value of the condition number for a random pattern shows a reasonably low condition number, and for the bunched pattern it is moderate, but they can influence the reconstructed signal depending on the noise level and error bounds.

The probability of generating a random pattern with a certain condition number is related to the distribution of the condition number that is discussed in [36]. The distribution of $cond(\mathbf{A}_C(k))$ for a typical case is shown in Fig.2.3. For example, the probability of taking a pattern with a condition less than 5 is in this case calculated to be 0.29.

Table I: various sample patterns and their condition numbers, $L=16, p=5$

| Type | $cond(\mathbf{A}_C(k))$ | C |
|---|---|---|
| **Optimal pattern** | **2.06** | **{2  3  9  10  14}** |
| Worst pattern | 8.35e+16 | {1  5  7  11  15} |
| Random pattern | 13.32 | {5  8  9  10  15} |
| Bunch pattern | 24.14 | {1  2  3  4  5} |
| **SFS pattern** | **2.06** | **{ 0  6  7  11  15}** |

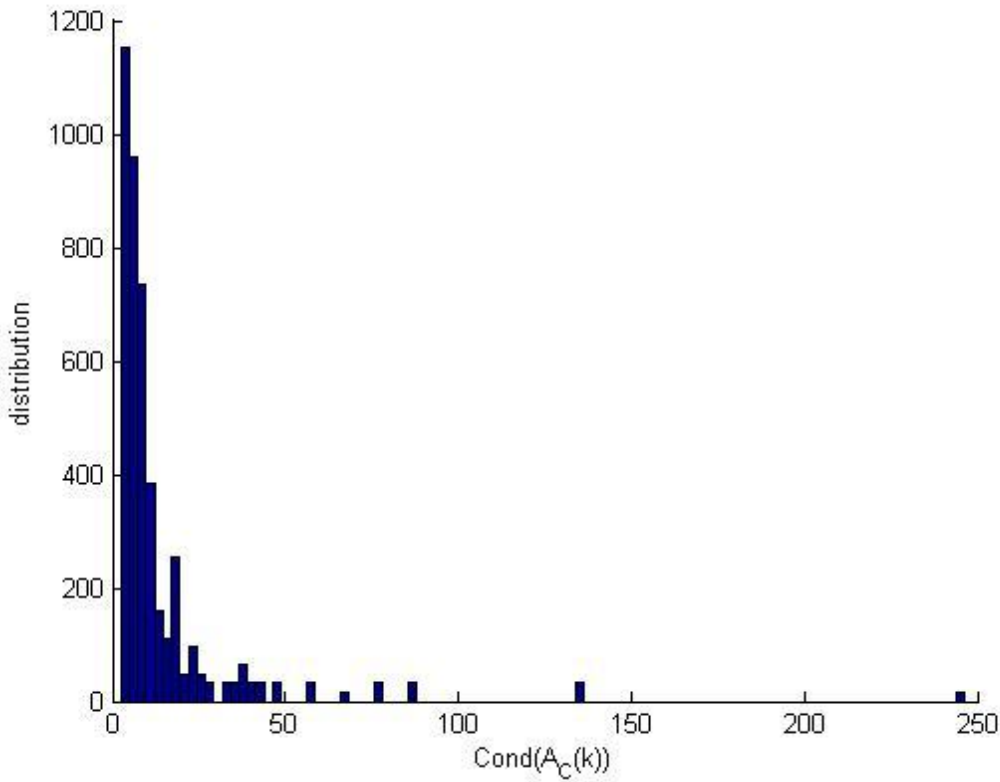

Fig.2.3: A typical distribution of the condition number, for $L=16$, $p=5$

Exhaustive search is infeasible for large values of $L$ and $p$; then we are looking for other search strategies to mitigate the cost of the search. The sequential search algorithms are practical for this desire. These algorithms add or remove features sequentially, but have a tendency to become trapped in local minima [13]. Representative examples of sequential search include sequential forward selection, sequential backward selection, plus-L minus-R selection, bidirectional search and sequential floating selection [13]. We used the sequential forward selection for choosing the sample pattern as follows.

**Sequential forward selection (SFS):**
Sequential forward selection is the simplest greedy search algorithm. Given a set $\mathbb{L} = \{0, 1, \ldots, L-1\}$, we want to find the subset $C=\{c_1, c_2, \ldots, c_p\}$, with $p < L$ that minimizes the objective function $cond(\mathbf{A}_C(k))$. It starts from the empty set and sequentially adds the feature $c^+$ that results in the minimum objective function when combined with the set $\boldsymbol{C}_i$ that have already been selected [13]. The algorithm for choosing the sample pattern with minimum condition number is summarized below:

1. Start with the empty set $\boldsymbol{C}_0 = \{\emptyset\}$
2. Select the next best future
$$c^+ = arg \min_{c \notin \boldsymbol{C}_i} [cond(\mathbf{A}_{\boldsymbol{C}_i \cup c}(\boldsymbol{k}))]$$
3. Update $\boldsymbol{C}_{i+1} = \boldsymbol{C}_i \cup c^+; i=i+1$
4. Go to step 2 if $i < p$

The search space is drawn like an ellipse to emphasize the fact that there are fewer states towards the full or empty sets. The main disadvantage of SFS is that it is unable to remove features that become obsolete after the addition of other features [13].

Nevertheless, the algorithm is easy to implement and much faster than exhaustive search. The total number of comparisons for choosing $p$ number out of $L$ in this way is derived with arithmetic progression as below:

The number of comparisons for the first element: *L*
The number of comparisons for the second element: *L-1*
.
.
.
The number of comparisons for the *p-th* element: *L-p+1*
Then total number of comparisons for the arithmetic progression is
$$S_p = pL - \frac{p(p-1)}{2}$$
(2.9)

For evaluation of the SFS algorithm we randomly generate *M=1000* spectral supports and find the corresponding SFS patterns and condition numbers and plot the histogram in Fig. 2.4. The result shows that the sample pattern achieved from the SFS search has a low condition number or sometimes the best one. Table I shows a SFS search where that the sample pattern gets the same condition number as the optimal one.

As an example when *L=32* and *p=10*, an exhaustive search needs $\binom{32}{10} = 64512240$ comparisons and with SFS search only 275 evaluations is needed. This shows a huge reduction in computations. The sample pattern derived in this approach for the spectral support of $\boldsymbol{k}=$ {3  4  6  7  12  13  18  19  21  22} is
$$\boldsymbol{C}=\{0 \quad 1 \quad 3 \quad 5 \quad 6 \quad 16 \quad 17 \quad 19 \quad 21 \quad 22\}$$

with $cond(\mathbf{A}_C(k))=2.8$, that is one of the best condition numbers.

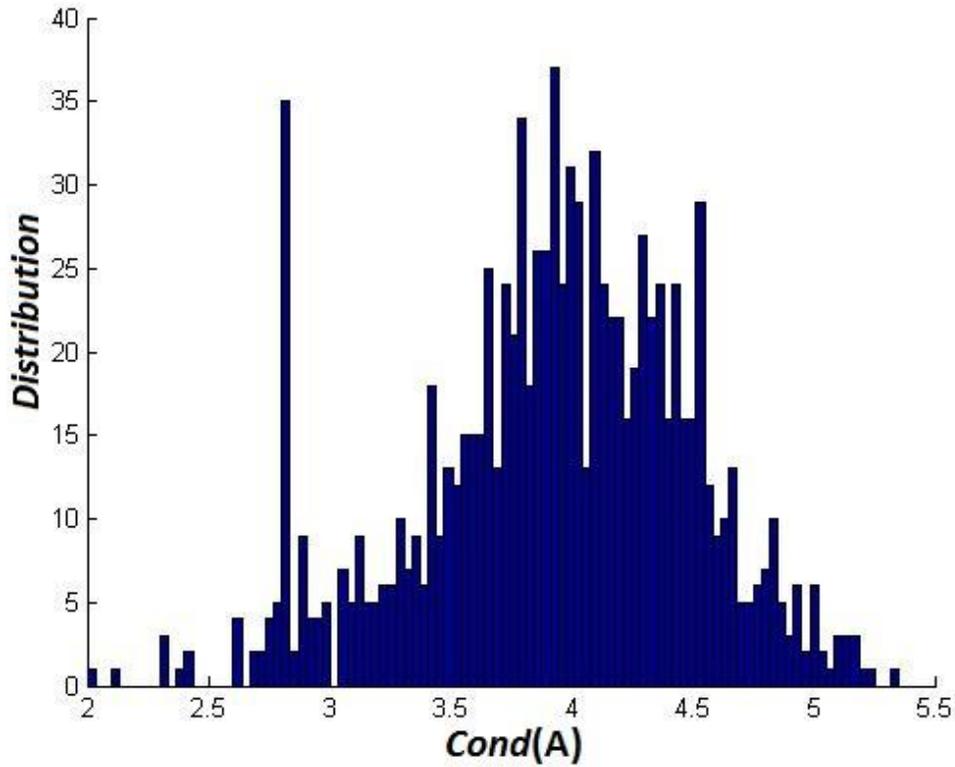

Fig.2.4: Distribution of the $Cond(\mathbf{A}_C(k))$ when sample pattern is selected with SFS algorithm

## 2.3 MATLAB simulation

The mentioned approach is used in a *MATLAB* simulation to sample and reconstruct a known spectrum multiband signal. The signal *x(t)* is generated as

$$x(t) = \sum_{i=1}^{N} a_i \ \text{sinc}(B_i(t-t_i))\exp(j2\pi f_i t)$$

(2.10)

where $\text{sinc}(x) = \sin(\pi x)/(\pi x)$ and *N*=3 is the number of bands. The *i*-th band of *x(t)* has width $B_i$, time offset $t_i$ and carrier frequency $f_i$. Fig.2.5 shows the time and frequency representations of the signal. The band locations are given as a set

$F$= {[0.7,1.3), [2.45, 2.75), [3.8 ,4.2)} and $f_{max}$=5.

The Lebesgue measure or Landauu lower bound for this signal is

$$\lambda(F) = \sum_{i=1}^{3} B_i = 0.6+0.3+0.4=1.3$$

The occupancy ratio is

$$\Omega = \lambda(F)/f_{max} = 1.3/5 = 0.26$$

The parameters *T*, *L*, *p* and set *C* should be selected to start the sampling. We set $T=1/f_{max}=0.2$ and arbitrarily $L=32$ as a moderate number. To set the parameter *p*, the spectral set should be derived based on (2.1) and (2.2) that results in

$$k=\{4\ \ 5\ \ 6\ \ 7\ \ 8,\ 15\ \ 16\ \ 17,\ 24\ \ 25\ \ 26\}$$

with $q=|k|=11$, so selecting $p=12$ is enough for a perfect reconstruction. The sampling pattern *C* that is obtained with the SFS search as

$$C=\{\ 0\ \ 1\ \ 2\ \ 6\ \ 11\ \ 12\ \ 13\ \ 18\ \ 22\ \ 23\ \ 24\ \ 28\}$$

that corresponding condition number with this sample pattern is $cond(A_C(k))=2.8$.

The average rate of sampling is $\bar{f} = \left(\frac{p}{L}\right) f_{max} = 0.375\, f_{max}$, that is close to Landau lower bound as $\Omega f_{max}= 0.26 f_{max}$.

The process of simulation generates $M=1024$ samples of $x(t)$ uniformly with $T=0.2$ according to (2.10), and then the sequence of $x_i[n]$ for $i=1,...12$ is created by picking the $c_i$-th sample and zero padding inter sample distance by $L$-$1=31$ zeros (1.11). The sequences are filtered with a low-pass filter with cut off frequency of $f_c=f_{max}/L$. We used the MATLAB command

$$h_r=\text{fircls1}(N_h,1/L,0.02,0.008)$$

to create a real-valued (non-ideal) low-pass filter $h_r[n]$ of length $N_h=383$, with normalized cut off frequency at $f_c=1/L$, pass-band ripple of 0.02 and stop-band ripple of 0.008, which is frequency shifted to obtain the complex filter $h[n] = h_r[n]\, exp(j\pi n/L)$ [8]. The operation of filtering with $h[n]$ introduces a delay of $t_d$ at the output that is equal to $t_d=(N_h+1)/2=192$ samples, hence the correct samples are started at sample number $t_d+1=193$.

Moreover, the matrix $A_C$ from (1.17) and then $A_C(k)$ is computed from (1.27), and then $A_C^\dagger$ is obtained by using the *MATLAB* command *"pinv"*. Finally, $x(nT)$ is reconstructed from (1.31). Fig.2.5 shows the original and reconstructed signal in the time and frequency domains. The simulation result is excellent. The relative reconstruction error defined as

$$RMSE = \frac{\|x_r[n] - x[n]\|}{\|x[n]\|}$$

(2.11)

is computed to be about 1.9%, while there is no non-ideality or noise.

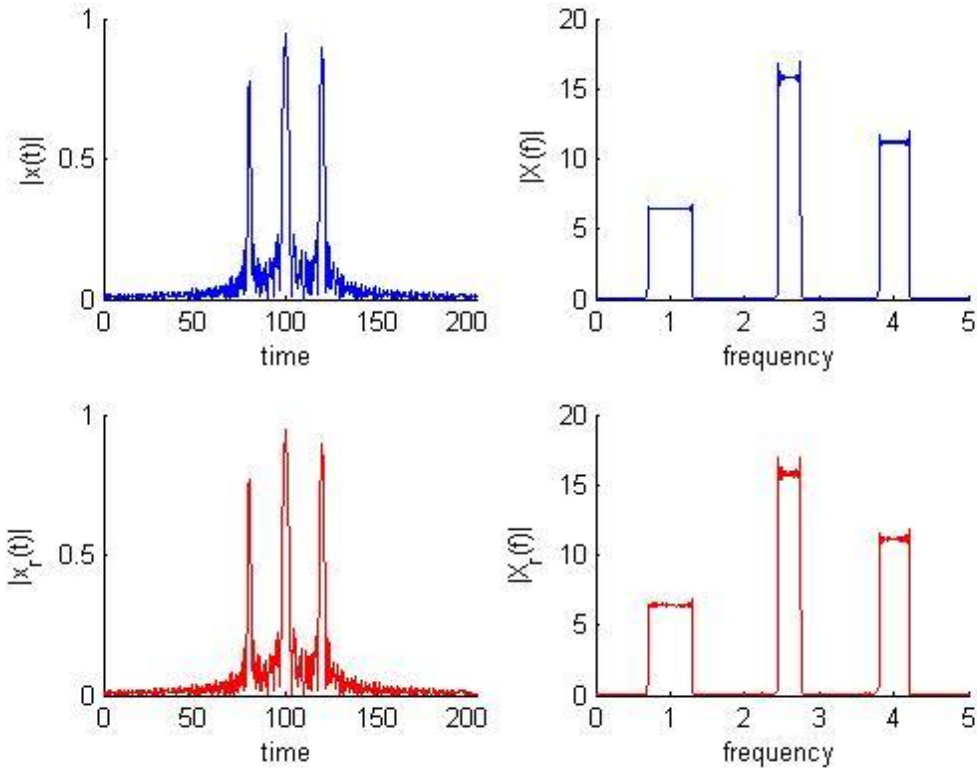

Fig.2.5: Input and reconstructed signal in the time and frequency domains. The relative reconstruction error is 1.9%.

## 2.4 Non-ideality effects

As non-ideality in uniform sampling is a limitation, here the reduced sampling rates afforded by non-uniform scheme can be accompanied by increased error sensitivity [3]. Non-idealities such as signal mismodeling, quantization error and jitter noise are sources of these errors. We model these non-idealities as an additive white sample noise; the sampled signal can be modeled as

$$\bar{x}(nT) = x(nT) + \Delta x(nT)$$

(2.12)

where $\Delta x(n)$ is the noise process with

$$E[\Delta x(mT)\, \Delta x(nT)] = \sigma^2\, \delta(m-n)$$

(2.13)

and *x(nT)* is the actual signal we would like to be sampling [3].
Owing to linearity, the output noise is derived from (1.28) as

$$\boldsymbol{y}(f) + \Delta \boldsymbol{y}(f) = \boldsymbol{A_C}(\boldsymbol{k})\, [\boldsymbol{z}(f) + \Delta \boldsymbol{z}(f)]$$

$$\Delta z(f) = A_C^\dagger \, \Delta y(f)$$

(2.14)

Therefore, the input error will be amplified by $A_C^\dagger$. From (2.7) the condition number and the inverse operation are directly related. Hence we appreciate the need for having a low condition number on the output noise.

To see the effect of non-ideality we repeat the above simulation with added quantization noise of an 8-bit ADC with input full range of $V_{FS}$=1.2V. According to (2.14), the level of noise at the output can vary depending on the sample pattern. Therefore, two distinct sample patterns with low and high condition numbers are used. The first one is the SFS sample pattern with a low condition number equal to 2.8 that, results a relative reconstruction error of 2.5%. The second one is a bunched sample pattern as

C={ 1    2    3    4    5    6    7    8    9    10    11 12}

with a condition number of 128, that results a relative reconstruction error of 36%. Fig. 2.6 depicts a comparison between the reconstruction error for both the cases of SFS search and the bunch pattern. Also, the reconstructed signal with quantization noise and the two different sample patterns are illustrated in Fig. 2.7.

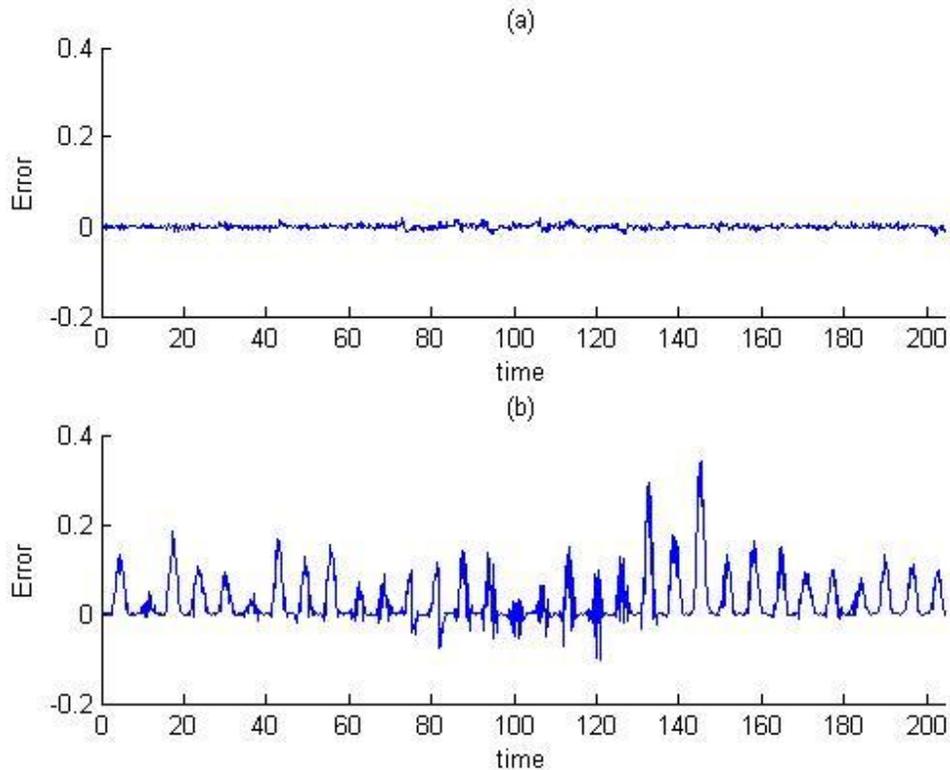

Fig.2.6: Output error in presence of quantization noise (a) SFS pattern, $cond(A_C(k))$=2.8 (b) bunched pattern and $cond(A_C(k))$=128.

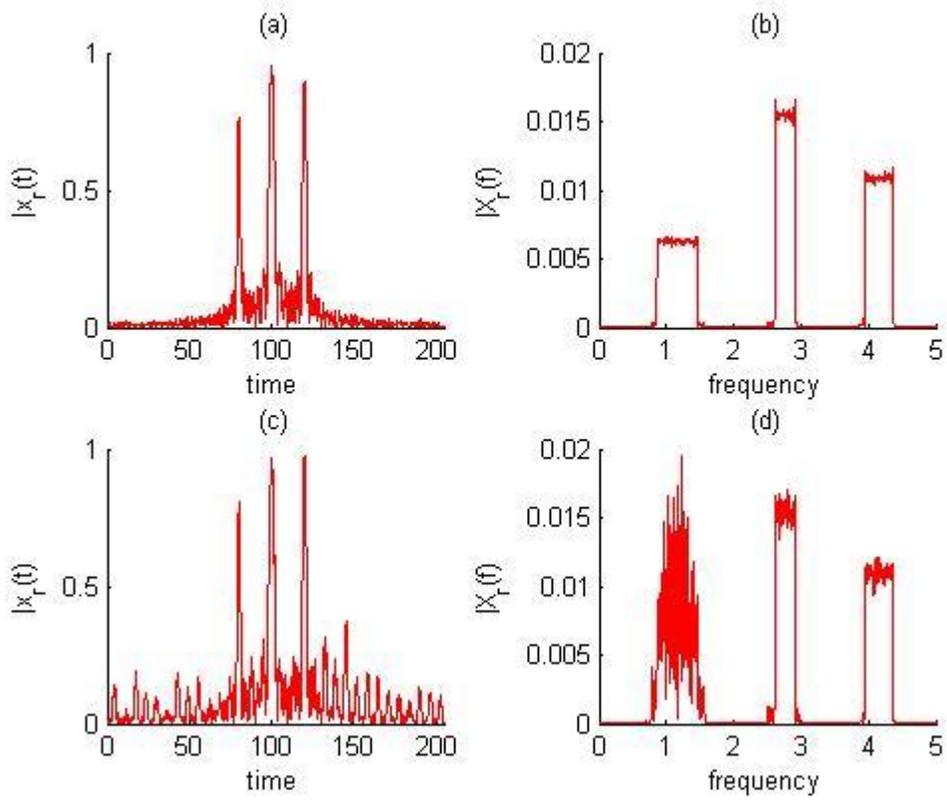

Fig. 2.7 Reconstructed signal in the time and frequency domains, in the presence of quantization noise (a),(b) SFS pattern, the relative error is 2.5% (c),(d) bunch pattern, the relative error is 36%

# 3. Unknown Spectral Support Signals

## 3.1 Introduction

If the information of the signal's spectral support is not available we encounter with the blind reconstruction problem. Actually, handling a minimum rate sampling and perfect reconstruction without any prior information of the signal spectrum is difficult. Therefore, we have to simplify our discussion with some assumptions about the signal to be sampled. Although these assumptions limit the discussion, they are not unrealistic.

In this way, in the case of unknown spectral support the band locations of the signals are not given as a set *F* such as in the previous case. However, we assume to know the number of bands, *N*, such that each band is no wider than *B*, and the maximal possible frequency of the signal $f_{max}$. Fig.3.1 shows a typical communication application that follows this structure. The values of *N, B, $f_{max}$* depend on the specifications of the application hand at [12]. In the example of Fig.3.1, *N*=3 and *B* is dictated by the widest transmission bandwidth. The multiband model does not assume knowledge of the carrier locations $f_i$, and these can lie anywhere below $f_{max}$.

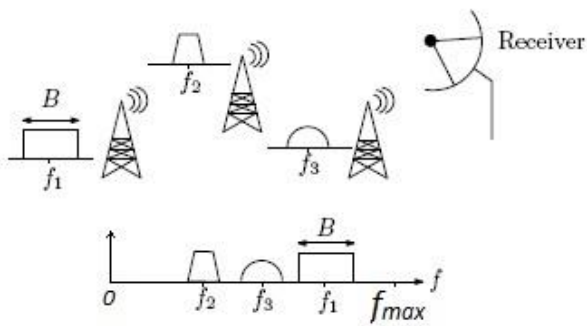

Fig.3.1: Three RF transmissions with different carriers $f_i$. The receiver observes a multiband signal [8].

## 3.2 Sampling Parameters

### 3.2.1 *L, p and q*

The selection of parameters *L*, *p* and *C* are more important in the unknown spectral case. In the first place the number of active slots is needed to discover but owing to unknown band locations it cannot be achieved exactly. With given *L* the number of active slots for the above model can be formulated as

$$\left\lceil \frac{NBL}{f_{max}} \right\rceil \leq q \leq N + N \left\lceil \frac{BL}{f_{max}} \right\rceil$$

(3.1)

where ⌈ ⌉ is the ceil function. Depending on the band locations the number of active slots can be any value between the two above bounds. For example in Fig. 3.2(a), the bands are such that they occupy minimum number of active cells that is $q_{min}$=3. While the band contents keep constant the carriers deviate such that they fill maximum number of active slots that is $q_{max}$=6 in Fig.3.2 (b).

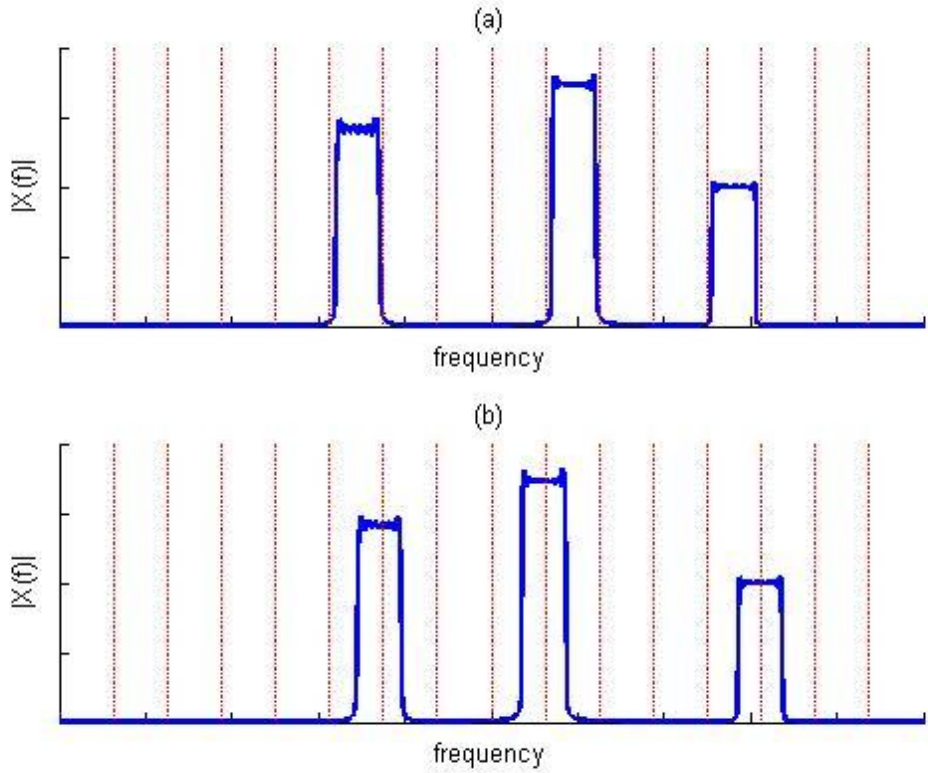

Fig. 3.2: The number of active slots changes with the band locations (a) Minimum number of active slots $q_{min}$ =3 (b) Maximum number of active slots $q_{max}$=6

According to this, the parameter $p$ is chosen such that is bigger than the maximum number of active slots

$$p > q_{max} = N + N \left\lceil \frac{BL}{f_{max}} \right\rceil$$

(3.2)

If choosing $L$ as

$$L = d \lfloor \frac{f_{max}}{B} \rfloor, \quad d \in \mathbb{N}$$

then

(3.3)

$$q_{max} = N + N \cdot d = N(d+1)$$

(3.4)

and

$$\bar{f} = \frac{p}{L} f_{max} = NB(1 + \frac{1}{d}) \approx NB$$

(3.5)

where $\Omega = \frac{NB}{f_{max}}$ is the occupancy of signal. Therefore by choosing $L$ according to (3.3) and a large $d$ the Landau lower bound can be achieved.

Although in the known spectrum case with choosing p ≥ q, the equation of (1.28) is solvable but in the case of unknown spectrum, depending on the approach that is chosen for recovery of spectral support we may need bigger values. This is discussed in Section 3.3.

### 3.2.2 Sample pattern

Choosing an appropriate sample pattern in the case of unknown spectral support is a bottleneck. The problem arises from the dependency of matrix $\mathbf{A}_C(\mathbf{k})$ on both sample pattern $C$ and spectral index set $\mathbf{k}$, as the rows of matrix are selected with $C$ and columns of matrix are selected with $\mathbf{k}$ elements (1.27). Then it may happen to find a suitable sample pattern for a specific spectral index set while it is worst when spectral index set changes. An optimal sample pattern can be found as the solution of the following min-max problem [9]

$$\boldsymbol{C}_{opt} = \arg \min_{C:|C|=p} \max_{\mathbf{k}:|\mathbf{k}|\leq p} cond(\mathbf{A}_C(\mathbf{k}))$$

(3.6)

where the symbol / / is the cardinality of the set or measure of the number of elements of the set.

This is a difficult combinatorial optimization problem, which is very likely NP-complete. Solution of (3.6) by exhaustive search would require $\binom{L}{p}^2$ evaluations of the condition number of $\mathbf{A}_C(\mathbf{k})$, which is infeasible for anything but the smallest problems [9]. To bypass the difficulty, the reference [9] suggests a combination of random search, heuristics and exhaustive testing. However, this approach is still feasible for small problems. Therefore, we use some heuristic tests and based on the observations give an algorithm to choose an optimal sample pattern that is more easily achieved and feasible.

### 3.2.2.1 Blind-SFS algorithm

First, we want to know how the condition number changes while the band location changes. For this desire, next scenario is applied: A suitable sample pattern is found by the SFS algorithm for a typical known spectral support. Next we move the band locations randomly and compute the condition number with the selected sample pattern and the new spectral support that is obtained after location movement. The histogram of these condition numbers is shown in Fig.3.3. Although the condition number changes after movement, still the result is not disruptive. This observation suggests that choosing an initial appropriate sample pattern by SFS algorithm reduces the dependency of $cond(\mathbf{A}_C(\mathbf{k}))$ on the $\mathbf{k}$, efficiently. It is noticeable that the value of condition number with an inappropriate sample pattern can reach infinity in this case.

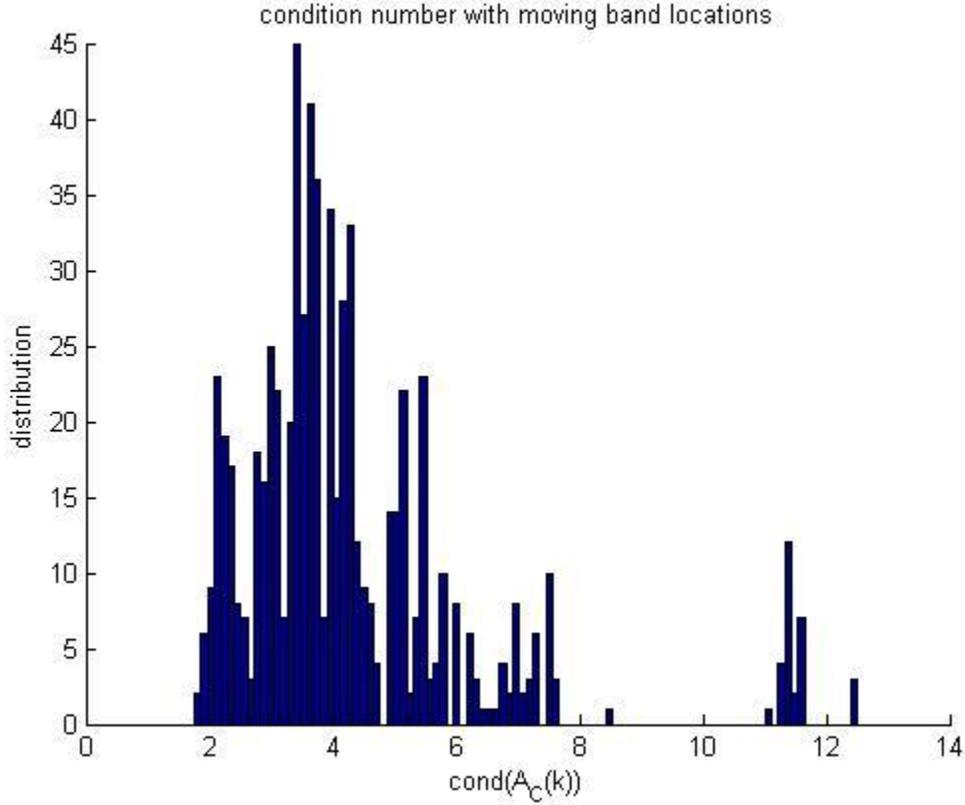

Fig. 3.3: Distribution of the condition number with varying band locations. The initial condition number is 2.2 and it may reach 13 after movement

Another interesting observation in this experiment is that as the distance between the bands reduces or they are overlapping the condition number increases. Using the fact that in practice we are facing with non overlaping bands and there are enough distance between carriers, we can be more hopeful to produce a general sample pattern with a low condition number.

On the other hand, given $N$, $B$, $f_{max}$ and computing the maximum and minimum number of active cells, the following simplification can be attained. Assume $N=3$, $B=1$, $L=f_{max}=20$. The number of active cells is in the range of $3 \leq q \leq 6$, therefore the spectral index set $k$ can be any of the following sets

$k1 = \{a, b, c\}$
$k2 = \{a, a+1, b, c\}$
$k3 = \{a, b, b+1, c\}$
$k4 = \{a, b, c, c+1\}$
$k5 = \{a, a+1, b, b+1, c\}$
$k6 = \{a, a+1, b, b+1, c, c+1\}$

the values of $a$, $b$ and $c$ are integers and such that the spectral index set $ki \subset \mathbb{L}$ for $1 \leq i \leq 6$.

In a testing scenario, the values of $a$, $b$ and $c$ are chosen uniformly and a sample pattern with the SFS search for the spectral index $k6$ is found. This sample pattern is then applied to the other possible spectral index sets $k1$ to $k5$. The condition numbers for all other spectral index set is then less than the condition number for $k6$, see Fig.3.4. In other words, if we choose a sample pattern that has a low condition number with the biggest possible spectral index set, here $k6$, it won't be worse for any other possible spectral index set that can be happen with the same parameters, here $k1$ to $k5$.

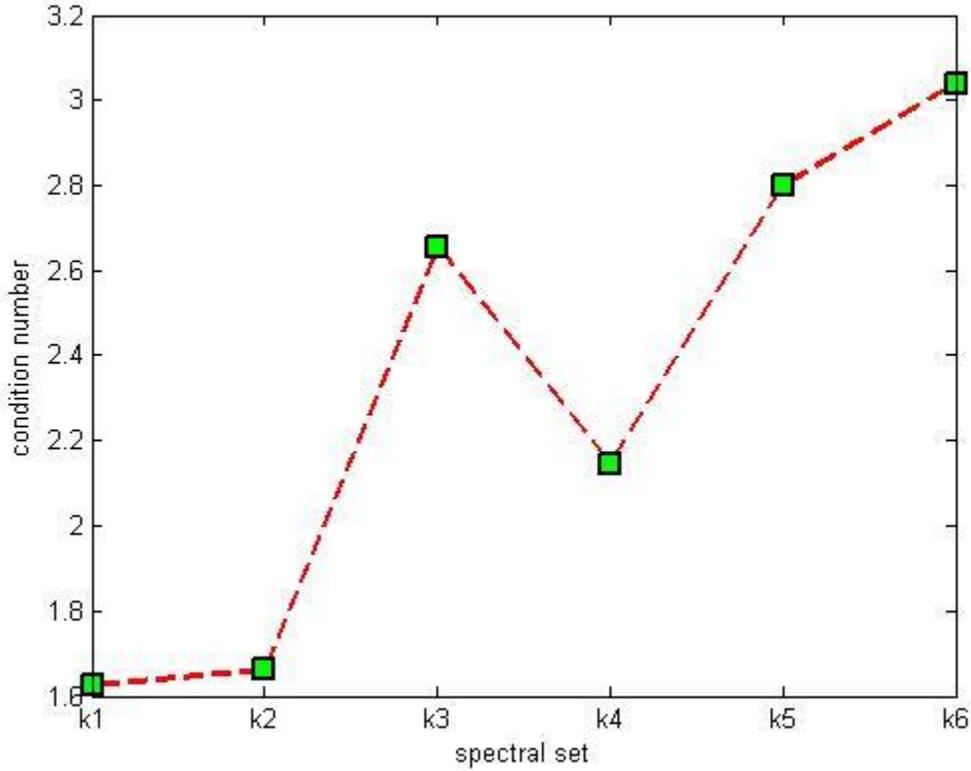

Fig 3.4: Sample pattern $C$ is designed for $k6$ but it works well for other spectral index set too.

Summarizing the results we give following instructions as a Blind-SFS algorithm for choosing the sample pattern in the case of unknown spectral support. Assume $N$, $B$ and $f_{max}$ are given:

1. Set $L = d \lfloor \frac{f_{max}}{B} \rfloor$
2. Compute $q_{max} = N(d+1)$ and set $p = q_{max}+1$
3. Set the spectral index set
   $k=\{a_1, a_1+1, ..., a_1+d, a_2, a_2+1, ..., a_2+d, ......, a_N, a_N+1, ..., a_N+d\}$
   where the coefficients $a_1, a_2, ..., a_N$ are selected uniformly random such that
   $$a_1+d < a_2,\ a_2+d < a_3,\ ...\ ,\ a_N+d < L \tag{3.7}$$
4. Select the sample pattern with SFS search with the derived parameters $L$, $p$ and $k$

The algorithm is fast even with large values of $L$, $p$ and $q_{max}$ and the results are reasonable. To evaluate the performance of the method we use it for a signal with $N=3$, $B=1.5$ and $f_{max}=20$. Using the algorithm

1. $d = 1$ and $L = \left\lfloor \frac{20}{1.5} \right\rfloor = 13$
2. $q_{max} = N(d+1) = 3*2 = 6$ and $p = 7$
3. Uniform random selection of $a_1 = 2, a_2 = 5, a_3 = 8 \Rightarrow k = \{ 2 \quad 3 \quad 5 \quad 6 \quad 8 \quad 9 \}$
4. The resulting sample pattern is $C = \{ 0 \quad 3 \quad 6 \quad 8 \quad 9 \quad 10 \quad 11 \}$

The selected sample pattern is applied to the signal with the same number of bands and different band locations that are chosen randomly. The histogram of condition numbers is shown in Fig.3.5. As the figure shows, the condition number is low and most of the time it is close to the desired value. However, we may get some moderate values, but hopefully these values are not disruptive and the probability of getting such values is small.

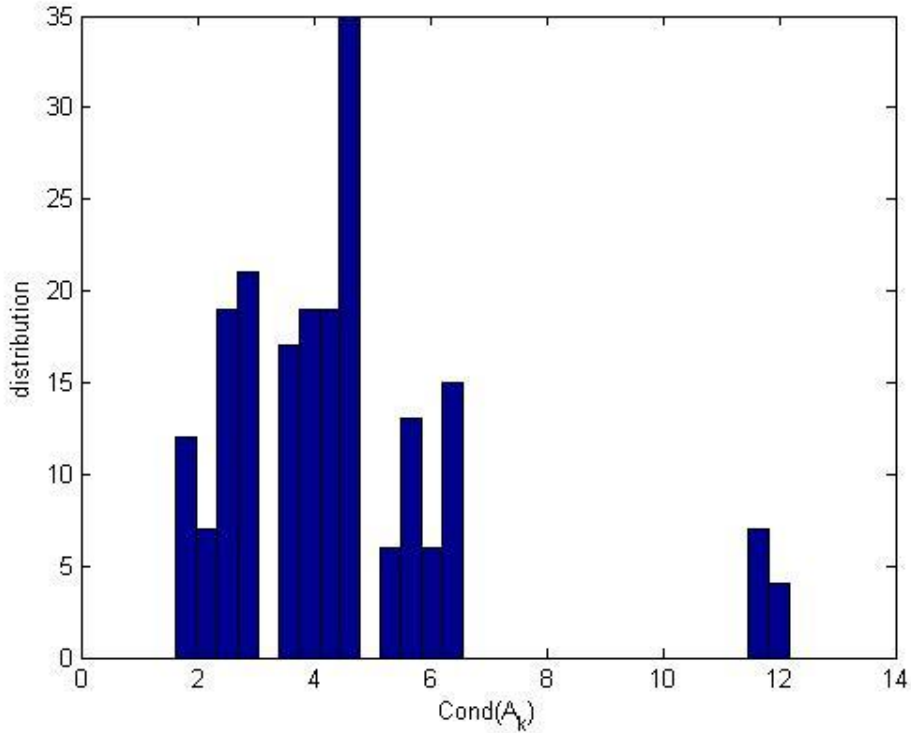

Fig 3.5: Selected sample pattern from the blind-SFS algorithm is applied to random signals and the distribution of the resulting condition number is displayed here.

## 3.3 Estimating the spectral index set

In contrast with the known spectrum case, where the spectral index set of the signal is computed easily using the band locations from (2.1) and (2.2), in the unknown spectrum case, the spectral index set should be determined based on the coset sampled data. In other words Equation (1.28) that is repeated here:

$$y(f) = A_C(k)z(f), \quad \forall f \in F_0$$

should be solved for finding both unknowns $k$ and $z(f)$. Therefore, given $y(f)$ the problem is to find the vector ***k*** with minimum length *q*, subject to (1.28) for some **z**(*f*) [5]. In a real situation there are some non-idealities that we can model by adding an additive noise vector of **n**(*f*) of size $p \times 1$ to (1.28) as

$$\mathbf{y}(f) = A_C(\mathbf{\textit{k}}) \mathbf{z}(f) + \mathbf{n}(f)$$

(3.8)

For simplicity, we assume that **n**(*f*) is a Gaussian complex noise with distribution $N(0,\sigma^2 \mathbf{I})$, which is also uncorrelated with the signal. Fortunately, this problem has the same form arising in Direction of Arrival (DOA) estimation in array processing and sinusoidal retrieval [5],[9]. To see the connections we briefly review the DOA model here.

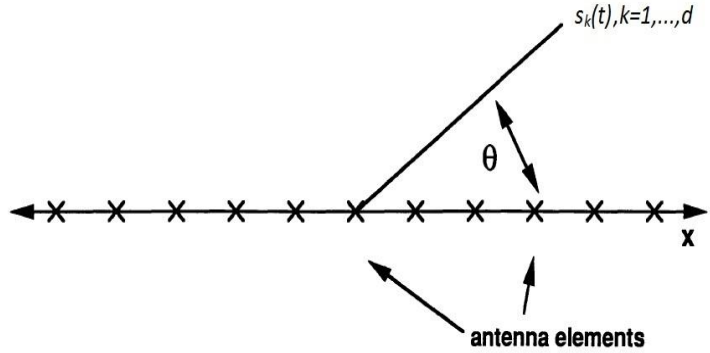

Fig.3.6: An antenna array system, one source and several antennas are indicated

In DOA estimation, *d* signal sources, $s_k(t)$, *k=1,…d*, are sampled by *m* antennas in different locations, see Fig.3.6. The received signals have different delays depending on the DOA [14], which are equivalent to phase shifts assuming narrowband signals . The model of received signals in matrix form can be expressed as

$$x(t) = \mathbf{A}(\boldsymbol{\theta}) \, s(t) + n(t)$$

(3.9)

Where $\boldsymbol{\theta}=[\theta_1,…,\theta_d]^T$ contains the signal parameters and $s(t)=[s_1(t),…,s_d(t)]^T$ is composed of the signal waveforms, $\mathbf{A}(\boldsymbol{\theta})$ is the steering matrix as

$$\mathbf{A}(\boldsymbol{\theta})_{m \times d} = \begin{bmatrix} e^{j\omega_c \tau_1(\theta_1)} & e^{j\omega_c \tau_1(\theta_2)} & \cdots & \cdots & e^{j\omega_c \tau_1(\theta_d)} \\ e^{j\omega_c \tau_2(\theta_1)} & e^{j\omega_c \tau_2(\theta_2)} & \cdots & \cdots & e^{j\omega_c \tau_2(\theta_d)} \\ \vdots & \vdots & \cdots & \cdots & \vdots \\ \vdots & \vdots & \cdots & \cdots & \vdots \\ e^{j\omega_c \tau_m(\theta_1)} & e^{j\omega_c \tau_m(\theta_2)} & \cdots & \cdots & e^{j\omega_c \tau_m(\theta_d)} \end{bmatrix}$$

(3.10)

Where $\omega_c$ is the carrier frequency and $\tau_k(\theta)$ denotes the propagation delay from the reference to the *k-th* element [14].

Compare (3.9) and (1.28) if *q=d* and *p=m*, reaches following equivalency:

$$\begin{bmatrix} z_1(f) \\ z_2(f) \\ \vdots \\ z_q(f) \end{bmatrix} \equiv \begin{bmatrix} s_1(t) \\ s_2(t) \\ \vdots \\ s_d(t) \end{bmatrix}, \quad \begin{bmatrix} y_1(f) \\ y_2(f) \\ \vdots \\ y_p(f) \end{bmatrix} \equiv \begin{bmatrix} x_1(t) \\ x_2(t) \\ \vdots \\ x_m(t) \end{bmatrix}$$

(3.11)

That shows that each active cell, $z_i(f)$, $i=1,...,q$, in (1.28) corresponds to a signal source, $s_k(t)$, in (3.9); and each coset sample sequence in the frequency domain $y_i(f)$, $i=1,...,p$, corresponds to an the output of an antenna $x_i(t)$. In other words, each active cell acts as a signal source located in the spectral index of $k_i \in \boldsymbol{k}$, $1 \leq i \leq q$, and generates the band frequency of $F_0 = [0, \frac{f_{max}}{L}]$. Also the steering matrix, $\mathbf{A}(\boldsymbol{\theta})$, and measurement matrix, $\mathbf{A_C}(\boldsymbol{k})$, have the same structure, the rows of $\mathbf{A}(\boldsymbol{\theta})$ are associated with the locations of sensors as the rows of $\mathbf{A_C}(\boldsymbol{k})$ are with the sample pattern, whereas the columns of $\mathbf{A}(\boldsymbol{\theta})$ are specified by angles $\boldsymbol{\theta}$ and the columns of $\mathbf{A_C}(\boldsymbol{k})$ by the spectral index set $\boldsymbol{k}$.

Assuming a flat signal in each cell, the quantity of SNR is directly related with the occupancy and amplitude of the signal in that cell. For illustration, consider Fig.3.7, where each signal band is divided into two slots with unequal occupancies, and they act as two different signal sources with unequal power. The cell with the bigger occupancy has the higher power. Modeling the blind system in this way takes the advantageous of having a good SNR even with low signal occupancy in each cell. This is because of reducing the noise power with a factor of $L$ in each cell.

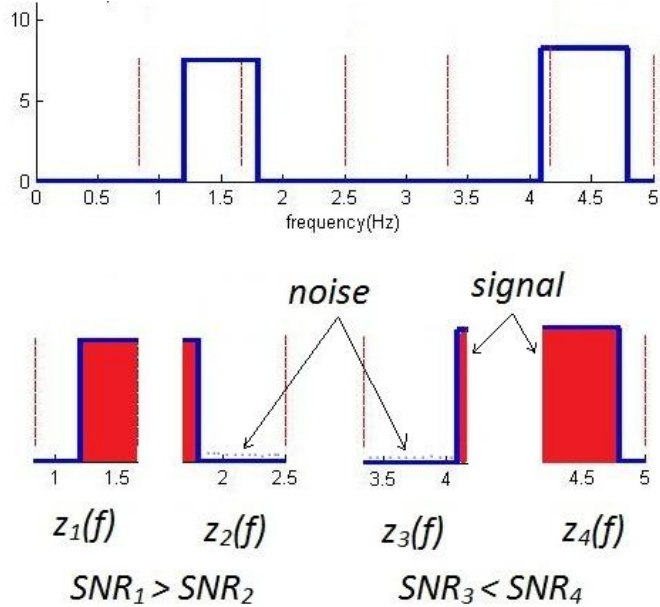

Fig.3.7: Sliced frequency representation of a wideband signal, each active cell acts as an independent source

Several approaches for solving this problem are suggested in the sensor array processing literature. All are based on the correlation matrix $\mathbf{R}$ defined as

$$\mathbf{R} = E[\mathbf{y}(f)\mathbf{y}^*(f)] = \mathbf{A}_C(\mathbf{k}) \mathbf{Z} \mathbf{A}_C^*(\mathbf{k}) + \sigma^2 \mathbf{I}$$

(3.12)

where ( )* denotes the Hermitian transpose, and $\mathbf{Z}$ is the correlation matrix of the signal vector $\mathbf{z}(f)$ [5],[14],[16] as

$$\mathbf{Z} = E[\mathbf{z}(f)\mathbf{z}^*(f)]$$

(3.13)

Depending on weather $\mathbf{Z}$ is full-rank or not, different methods would be applied. We consider some of these methods in the next section.

### 3.3.1 Subspace methods

If $\mathbf{Z}$ is full rank of $q$, that means the signal vector $\mathbf{z}(f)$ are not coherent, the geometrical properties of the correlation matrix can be used. From (3.12) it can be seen that any vector that is orthogonal to $\mathbf{A}_C(\mathbf{k})$ is an eigenvector of $\mathbf{R}$ with corresponding eigenvelue $\sigma^2$. The remaining eigenvectors are all in the range space of $\mathbf{A}_C(\mathbf{k})$, and are therefore termed signal eigenvectors. The eigen-decomposition of $\mathbf{R}$ is partitioned into a signal and a noise subspace as [14]

$$\mathbf{R} = \sum_{k=1}^{p} \lambda_k \mathbf{e}_k \mathbf{e}_k^* = \sum_{k=1}^{q} \lambda_k \mathbf{e}_k \mathbf{e}_k^* + \sum_{k=q+1}^{p} \lambda_k \mathbf{e}_k \mathbf{e}_k^*$$
$$= \mathbf{E}_s \mathbf{\Lambda}_s \mathbf{E}_s^* + \mathbf{E}_n \mathbf{\Lambda}_n \mathbf{E}_n^*$$

(3.14)

where

$$\mathbf{\Lambda}_{S_{q \times q}} = \begin{bmatrix} \lambda_1 & 0 & \cdots & 0 \\ 0 & \lambda_1 & \cdots & 0 \\ \vdots & \vdots & \vdots & \vdots \\ 0 & 0 & . & \lambda_q \end{bmatrix}, \mathbf{\Lambda}_{n_{(p-q) \times q}} = \begin{bmatrix} \lambda_{q+1} & 0 & \cdots & 0 \\ \vdots & \vdots & \vdots & \vdots \\ 0 & 0 & . & \lambda_p \end{bmatrix}$$

(3.15)

here $\lambda_1 \geq \lambda_2 \geq \cdots \lambda_q > 0$, are the signal eigenvalues and $\mathbf{E}_s = [\mathbf{e}_1, \ldots, \mathbf{e}_q]$ is the matrix of the corresponding $q$ eigenvectors. Further $\lambda_{q+1} = \cdots = \lambda_p = \sigma^2$, are noise eigenvalues and $\mathbf{E}_n = [\mathbf{e}_{q+1}, \ldots, \mathbf{e}_p]$ is the matrix of the corresponding ($p$-$q$) noise eigenvectors [14],[5]. The signal eigenvectors in $\mathbf{E}_s$ span the range space of $\mathbf{A}_C(\mathbf{k})$, which is termed the signal subspace [14]. For the noise eigenvector we have instead, $\mathbf{E}_n \perp \mathbf{A}_C(\mathbf{k})$. Then, from the spectrum of $\mathbf{R}$ with eigenvalues in decreasing order, it becomes easy to discriminate between signal and noise eigenvalues, and hence determination of the number of active slots $q$ would be attained [16],[14],[5]. In this way the first step in the subspace methods is to find the number of signal eigenvalues. This issue is underlying the model order selection problem and will be considered later.

As the distribution of signal is unknown the real correlation matrix $\boldsymbol{R}$ cannot be achieved. Then $\boldsymbol{R}$ is estimated from the measured data as

$$\boldsymbol{R} \triangleq \int_0^{\frac{f_{max}}{L}} \boldsymbol{y}(f)\, \boldsymbol{y}^*(f) df$$

(3.16)

with the dimension of $p \times p$, the ( )* denotes the Hermitian transpose [5]. Substituting $y(f)$ from (3.8) to (3.16), we have [5]

$$\boldsymbol{R} = \int_0^{\frac{f_{max}}{L}} \mathbf{A}_C(\boldsymbol{k})\, \boldsymbol{z}(f)\, \boldsymbol{z}^*(f)\, \mathbf{A}^*{}_C(\boldsymbol{k})\, df$$

$$= \mathbf{A}_C(\boldsymbol{k}) \int_0^{\frac{f_{max}}{L}} \boldsymbol{z}(f)\, \boldsymbol{z}^*(f)\, df\ \mathbf{A}_C^*(\boldsymbol{k})$$

$$= \mathbf{A}_C(\boldsymbol{k})\, \mathbf{Z}\, \mathbf{A}_C^*(\boldsymbol{k})$$

(3.17)

where $\mathbf{Z} \geq 0$ is a $q \times q$ matrix given by

$$\mathbf{Z} = \int_0^{\frac{f_{max}}{L}} \boldsymbol{z}(f) \boldsymbol{z}^*(f) df$$

(3.18)

$\mathbf{Z}$ is a Gram matrix of the functions $z_r, r = 1, \ldots, q$ defined in (1.26). It follows that $\mathbf{Z}$ has full rank if these functions are linearly independent [5].

From Parseval's identity the correlation matrix can be computed directly in the time domain from the filtered sequences $x_{hi}[n]$ in (1.30) using the formula [5],[8],[9]

$$(\boldsymbol{R})_{kl} = \langle x_{hk}[n], x_{hl}[n] \rangle = \sum_{n=-\infty}^{n=+\infty} x_{hl}[n]\, x_{hk}^*[n]$$

(3.19)

where $\langle\,\rangle$ denotes the inner product operation. Therefore the computation cost is linear in the amount of data [5]. In practice the number of samples is limited and therefore a sample correlation matrix is defined based on $M$ available samples as

$$\left(\widehat{\boldsymbol{R}}\right)_{kl} = \frac{1}{M} \sum_{n=1}^{n=M} x_{hl}[n]\, x_{hk}^*[n]$$

(3.20)

Under suitable assumptions $\widehat{\mathbf{R}} \to \mathbf{R}$ when $M \to \infty$.

3.3.1.1 **Estimating the number of active slots**

As we mentioned before the number of active slots is the order of the model in (3.8) that can be estimated from ordered eigenvalues of sample correlation matrix. Suppose the $p$ sequences $x_{hi}[n]$, $i=1,...,p$ according to (1.30) are provided. The sample correlation matrix $\widehat{R}$ is computed from $M$ samples according to (3.20), and the eigendecomposition will be

$$\widehat{R} = \widehat{E}_s \widehat{\Lambda}_s \widehat{E}_s^* + \widehat{E}_n \widehat{\Lambda}_n \widehat{E}_n^*$$

(3.21)

The $p$ ordered eigenvalues are as follows

$$\lambda_1 \geq \cdots \lambda_q \cdots \geq \lambda_p$$

where $q$ eigenvalues are significant and $(p-q)$ eigenvalues are ideally in the range of the noise. Fig. 3.8 depicts a typical case of $p=10$ ordered eigenvalues, with seven significant and three small eigenvalues. As we see, there is a gap between $\lambda_7$ and $\lambda_8$ that depends on the SNR, and the total number of eigenvalus. Therefore, choosing $q$ significant eigenvalues out of $p$ needs the subjective judgment in selecting the threshold levels for the different tests [17].

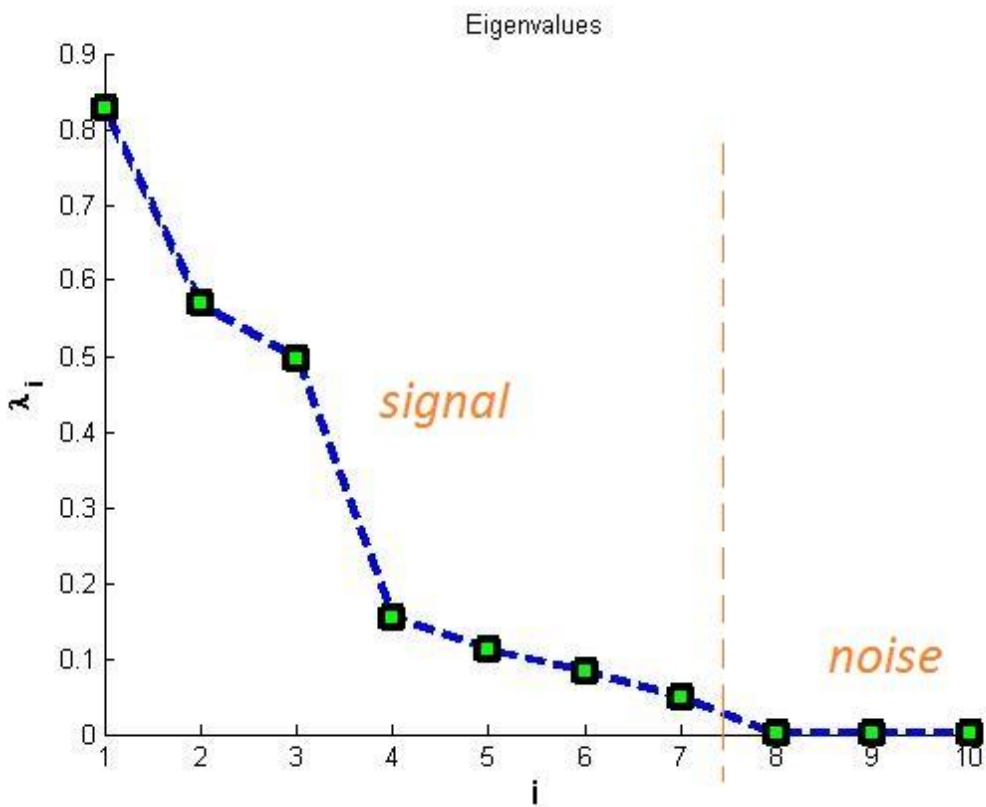

Fig. 3.8: Typical ordered eigenvalues. Note that there is a gap between the signal and noise eigenvalues that should be detected by subjective judgment, $p=10$ and $q=7$

Information theoretic criteria (ITC) approaches have been widely suggested for this kind of problem. The best known of this test family are the Akaike information criterion (AIC) and the minimum description length (MDL) [16]. The number of active cells is determined as the value for which the AIC or the MDL criteria is minimized [18]. The number of active slots using the AIC is the integer $\hat{q}$ which satisfies [16][17][18]:

$$\hat{q} = \arg\min_{r} -M(p-q)\log\left(\frac{g(r)}{a(r)}\right) + q(2p-r), \quad q_{min} \leq r \leq q_{max}$$

(3.22)

Here $M$ is the number of samples, $g(r)$ is the geometric mean of the eigenvalues [18]

$$g(r) = \prod_{i=r+1}^{p} \lambda_i^{\frac{1}{p-r}}$$

(3.23)

and $a(r)$ is the arithmetic mean of the eigenvalues [18]

$$a(r) = \frac{1}{p-r} \sum_{i=r+1}^{p} \lambda_i$$

(3.24)

The number of active slots using the MDL criterion is given by

$$\hat{q} = \arg\min_{r} -M(p-r)\log\left(\frac{g(r)}{a(r)}\right) + \frac{1}{2}r(2p-r)\log M$$

(3.25)

Above, we have assumed that the noise samples are white. But the samples involved in computing the correlation matrix are filtered by the interpolating filter (*h*), and then the corresponding noise samples may be correlated after filtering. However, the correlation among the noise samples is only related to the interpolating filter. Thus, the noise correlation matrix can be found based on the interpolating filter, and pre-whitening techniques can be used to whiten the noise samples [19].

The probability of correct detection of number of active slots, that is $P_d = Prob[\hat{q} = q]$, depends on the number of samples *M*, *SNR* and the noise distribution, and it is not equal to one all the time. But it could be enhanced by using the operation of peak detection which is used in MUSIC detection. Fig.3.9 illustrates the detection probability of the number of active slots for a typical signal. It is seen that MDL has a better performance than AIC in this case.

The article [16] introduces another technique of model order selection called exponential fitting test (EFT) that is claimed to be effective for short data. This method exploits the exponential profile of the ordered noise eigenvalues. For white Gaussian noise and short data it is shown that the profile of the ordered noise eigenvalues is seen to approximately fit an exponential law [16]. Assuming that the smallest eigenvalue is the noise eigenvalue, this exponential profile can then be used to find the theoretical profile of the noise-only eigenvalues. Starting with the smallest eigenvalue a recursive algorithm is then applied in order to detect a mismatch greater than a threshold value between each observed eigenvalue and the corresponding theoretical eigenvalue. The occurrence of such a mismatch indicates the presence of a source, and the eigenvalue index where this mismatch first occurs is equal to the number of sources present [16].

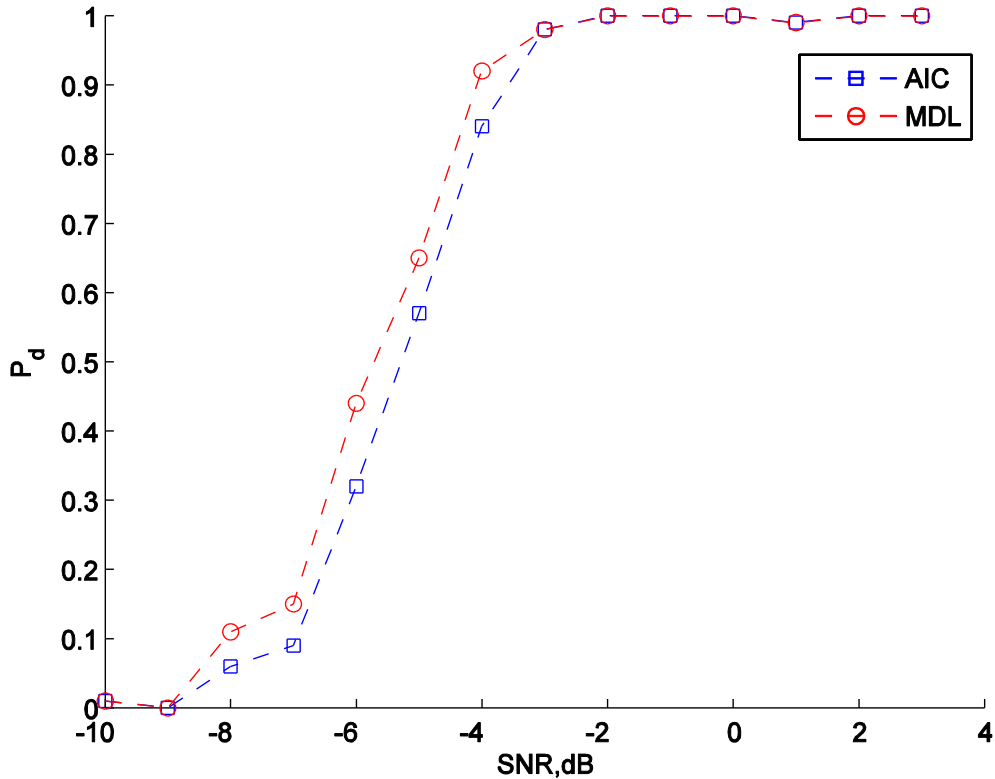

Fig. 3.9: Detection probability of number of active slots for a typical multi-band signal

The profile of the theoretical noise only eigenvalues is compared with the profile of the signal in presence of white additive noise in Fig. 3.10. In the case of noise only, the profile keeps the exponential form, while the profile of the signal with noise starts deviating from the exponential form at $\lambda_7$. The main idea of the test is to detect the eigenvalue index at which a break occurs between the profile of the observed eigenvalues and the theoretical noise eigenvalue profile provided by the exponential model [16]. The break point is detected by comparing the relative difference between the theoretical noise eigenvalue and the observed eigenvalue with a determined threshold. Fig. 3.10 shows the relative difference for the profile of Fig. 3.9. As expected the difference becomes small when index reaches $i=7$, which suggests seven significant and three noise eigenvalues in the spectrum of the signal.

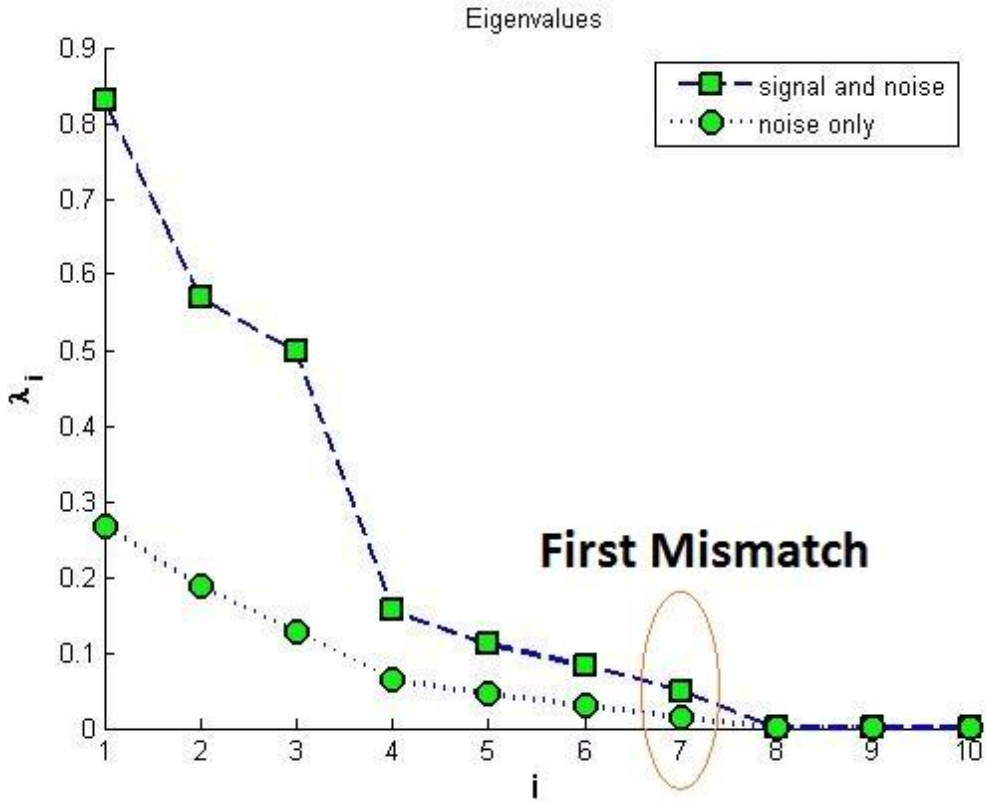

Fig. 3.10: Profile of the ordered noise and signal eigenvalues. The first mismatch occurs at index *i=7*

### 3.3.1.2 **Location of active slots using a MUSIC-Like algorithm**

After estimation of the number of active slots, $\hat{q}$, the location of the active slots can be recovered according to a MUSIC-Like algorithm as

$$P_{MU}(k) = \frac{\|a(k)\|^2}{\|a^*(k)\widehat{E}_n\|^2}, \qquad 0 \leq k \leq L-1$$

(3.26)

where *k* is the spectral index and *a(k)* is the *k-th* column of $\mathbf{A_C}$, given by

$$a(k) = \frac{1}{LT}\begin{bmatrix} e^{\frac{j2\pi k\, c_1}{L}} \\ e^{\frac{j2\pi k\, c_2}{L}} \\ \vdots \\ e^{\frac{j2\pi k\, c_p}{L}} \end{bmatrix}$$

(3.27)

The relation (3.26) will generate $L$ values for $L$ spectral indices such that if $k$ is an active cell, the value of $P_{MU}$ is significant in that point, and otherwise it will be smaller than a threshold. The location of the active slots is then specified by choosing $\hat{q}$ significant values of the computed $P_{MU}$:

$$\widehat{\boldsymbol{k}} = \{k_i | P_{MU}(k) > threshold\}$$

(3.28)

Fig.3.11 depicts the computed values of $P_{MU}$ for a typical signal with $\hat{q} = 7$ and $L=32$. As seen in the figure there are seven significant values and their locations specify the spectral index set of the signal, that is

$$\widehat{\boldsymbol{k}} = \{4,5,11,12,13,24,25\}$$

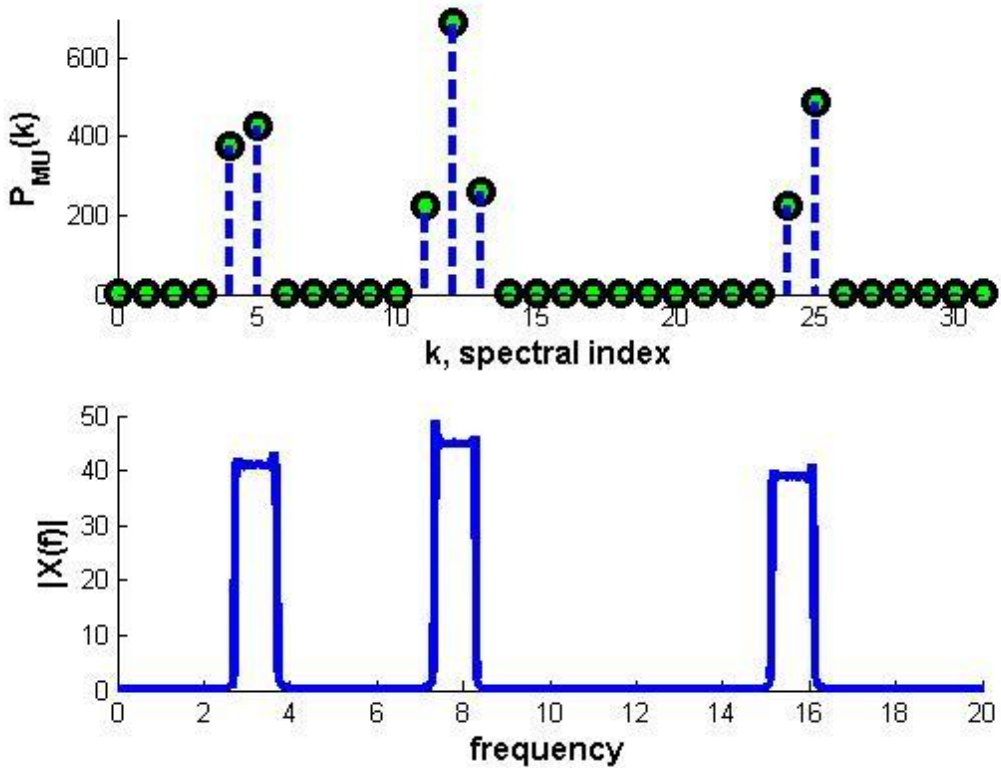

Fig. 3.11: Spectrum and the $P_{MU}$ values of a typical 3-bands signal. The locations of the significant values specify the spectral index set of the signal, $L=32$, $\hat{q} = 7$, $\boldsymbol{k}=\{4,5,11,12,13,24,25\}$

### 3.3.2 Least squares-based spectral estimation
In case of coherent signals, the matrix $\mathbf{Z}$ is not full rank anymore. Consider the model (1.28) again $\boldsymbol{y}(f) = \mathbf{A}_C(\boldsymbol{k})\,\boldsymbol{z}(f)$. The problem of finding a spectral index $\boldsymbol{k}$ with $q$ elements for some signals $\mathbf{z}(f)$ can be solved by using a Non-Linear Least-Squares (NLLS) approach as [14]

$$\{\hat{\mathbf{k}}, \hat{\mathbf{z}}(f)\} = \arg \min_{k,z(f)} \int_0^{\frac{f_{max}}{L}} \|\mathbf{y}(f) - \mathbf{A}_C(\mathbf{k})\mathbf{z}(f)\|^2 \, df$$

(3.29)

This is a separable least-squares problem, and for fixed (but unknown) $\mathbf{k}$, the solution with respect to the linear parameter $\mathbf{z}(f)$ is [14]

$$\hat{\mathbf{z}}(f) = \mathbf{A}_C^\dagger \, \mathbf{y}(f)$$

(3.30)

Substituting (3.30) into (3.29) leads to the concentrated NLLS formulation

$$\hat{\mathbf{k}} = \arg \min_k Tr\{\mathbf{P}_k^\perp \, \hat{\mathbf{R}}\}$$

(3.31)

with

$$LS(\mathbf{k}) = Tr\{\mathbf{P}_k^\perp \, \hat{\mathbf{R}}\}$$

(3.32)

The above interpreted as the power error between the measurements data and the estimated signal that should be minimized for a correct estimation, and

$$\mathbf{P}_k^\perp = \mathbf{I} - \mathbf{P}_k = \mathbf{I} - \mathbf{A}_C(\mathbf{k})\mathbf{A}_C^\dagger$$

(3.33)

is the orthogonal projection onto the nullspace of $\mathbf{A}_C^*(\mathbf{k})$ [14],[5].

As the exhaustive search for solving (3.31) needs choosing $q$ active cells out of $L$, that is solvable only for small $q$ and $L$ [5]. A practical approach at a reasonable cost is to employ a sequential search where one cell of the spectral index is selected at the time to minimize the criterion in (3.31) [14]. The procedure is the same as mentioned in section 2.2.3.1, but with a different target function this time. It starts from the empty set and sequentially adds the cell that produces the minimum value of the criterion in (3.31). As the exact number of active cells $q$ is unknown, the sequential selection should be repeated $q_{max}$ times. The total number of searches in this way as calculated in (2.9) is less than ($L \, q_{max}$). Meanwhile, the correct active cell is augmented to the set; the value of least square criterion diminishes monotonically and it becomes zero at the perfect estimation point. After this point, adding any other cell does not reduce the criterion. Fig.3.12 shows a typical result of the least square operation for a signal with six active cells and $q_{max}=9$. As seen, the error decreases rapidly toward $q=6$ and after that it becomes constantly a small value close to zero. Therefore, with choosing a threshold, $\epsilon$, as a reasonable residual error based on noise model, the procedure of search can be shortened before $q_{max}$. In summary, the algorithm is as below

1. Start with the empty set $\mathbf{k}_i = \{\emptyset\}$
2. Select the next cell such that $\quad k^+ = \arg \min_{k \notin k_i} LS(\mathbf{k}_i \cup k)$

3- Update $k_{i+1} = k_i \cup k^+$; $i=i+1$
4- Go to step 2 if $i < q_{max}$ or $LS(k_{i+1}) > \epsilon$

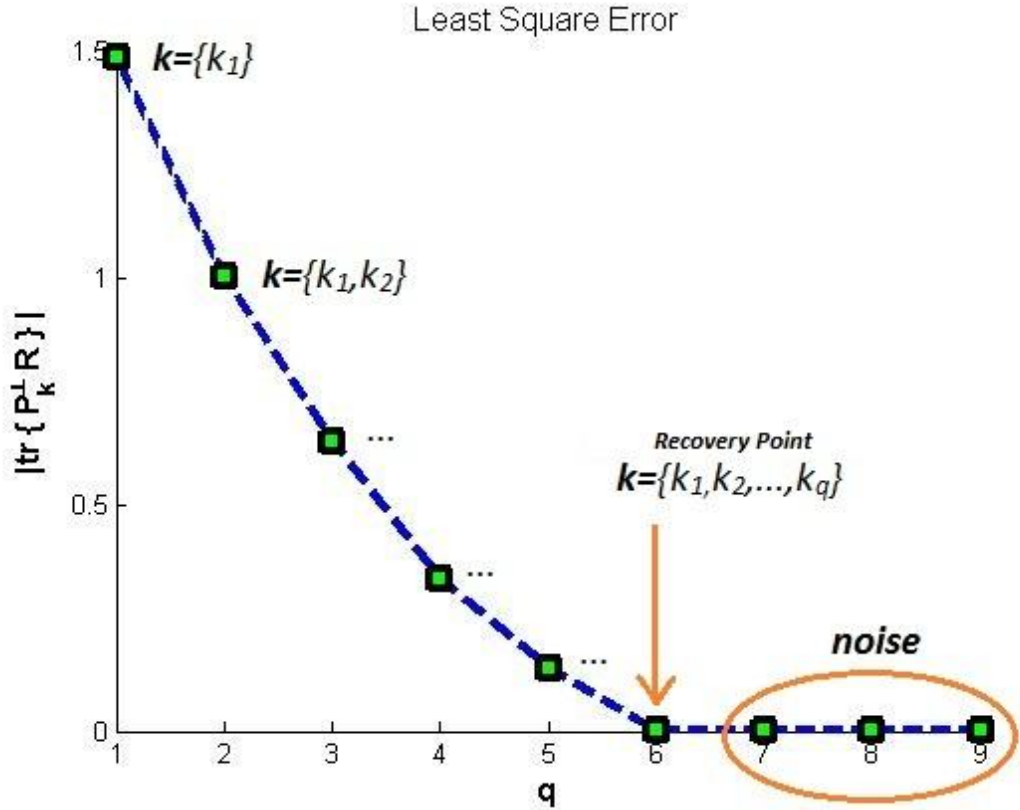

Fig.3.12: The Least square error monotonically decreases with $q$ and $k$

In contrast with independent signals, when choosing $p \geq q+1$ is enough for a perfect recovery of the spectral index set and signal; if **Z** is not full rank we need to choose a higher value to compensate for this deficiency. In the article [5],[8], it is proved that choosing $p \geq 2q$ guarantees the recovery even in the worst case where rank(**Z**)=1. As $q$ is unknown, we have to choose $p \geq 2q_{max}$.

In summary, rank deficiency of **Z** or linear dependency imposes a very special restriction on $x(t)$, in addition to its spectral sparsity [5] that is almost impossible to meet in practical applications. Therefore, we can assume the matrix **Z** is always full rank, and the subspace method can be used for almost all signals, although the least-square algorithm works for all signals, regardless of rank of **Z**, or equivalently, of the shape of the spectrum of the signal over its support [5].

### 3.3-3 Other methods:

In the article [8] two other methods are suggested to find the spectral index set that we mention here:

The naive approach for solving the equation (1.28) is to discretize the frequency interval $F_0$ to an equally spaced finite grid $\{f_i\}_{i=1}^{G}$ and then solve the equation only for $\mathbf{z}(f_i)$. The resulting finite dimensional problem can be solved within the regular compressed sensing framework [8].

The other approach is again based on the correlation matrix. It changes the problem into a problem of Multiple Measurement Vector (MMV) and then uses the solution of the MMV system from compressed sensing. The algorithm is given as:
1- Compute the sample correlation matrix **R** from (3.20)
2- Decompose $\mathbf{R} = \mathbf{V}\mathbf{V}^H$, where **V** is a $p \times r$ matrix, where $r = rank(\mathbf{R})$
3- Solve the linear system $\mathbf{V} = \mathbf{A}_C \mathbf{U}$ for the sparset solution $\mathbf{U}_0$
4- The spectral index set is then $k = \mathbf{I}(\mathbf{U}_0)$, where $\mathbf{I}(\mathbf{U}_0)$, the support of $\mathbf{U}_0$ indicates the rows of $\mathbf{U}_0$ that are non-identically zero.

The algorithm needs to find the sparsest solution $\mathbf{U}_0$ in the third step, which is known to be an NP-hard problem [8]. The MMV solvers such as the brute-force method, multi-orthogonal matching pursuit (M-OMP) are addressed in the compressed sensing literature [20]-[24] for finding $\mathbf{U}_0$ [8].

### 3.4 MATLAB Simulation

The process of blind spectrum sampling and reconstruction is implemented in *MATLAB* and presented here. The signal is generated according to the model of (2.10) with $N=3$ bands, $f_i=[4.8,10.45,15.4]$, $B_i=0.9$, $t_i=[6,13,17]$ and $f_{max}=20$. The $N$, $B_i$ and $f_{max}$ are assumed to be known. The parameters are selected as follows

1- From (3.3), $d=1$ and $L = \left\lfloor \frac{fmax}{B_i} \right\rfloor = \left\lfloor \frac{20}{0.9} \right\rfloor = 22$
2- From (3.4), $q_{max} = N(d+1) = 6$, $p = q_{max}+1 = 7$
3- Choosing sample pattern using the Blind-SFS algorithm results in
   C= {0  5  6  8  11  16  17}, with $cond(\mathbf{A}_C(k)) = 2$
4- Compute the sample correlation matrix $\widehat{\mathbf{R}}_{7\times 7}$ from (3.20)
5- Compute the eigenvalues $\lambda_1$ to $\lambda_7$ and eigenvectors of **R**. Tthe plot of the ordered eigenvalues is shown in Fig.3.13
6- From (3.22) or (3.25), estimate the number of active cells, which gives $\hat{q} = 5$
7- Find the noise eigenvectors $\mathbf{U}_n = [\mathbf{e}_6, \mathbf{e}_7]$ from (3.14) and (3.15)
8- The estimated spectral index from (3.26) and (3.28) is $\widehat{k} = [4\ 5\ 11\ 16\ 17]$
9- After finding the spectral index, the procedure of reconstruction is the same as for the case of a known spectrum

Fig.3.14 illustrates the input and the reconstructed signal, the relative *MSE* error is around 2.7%.

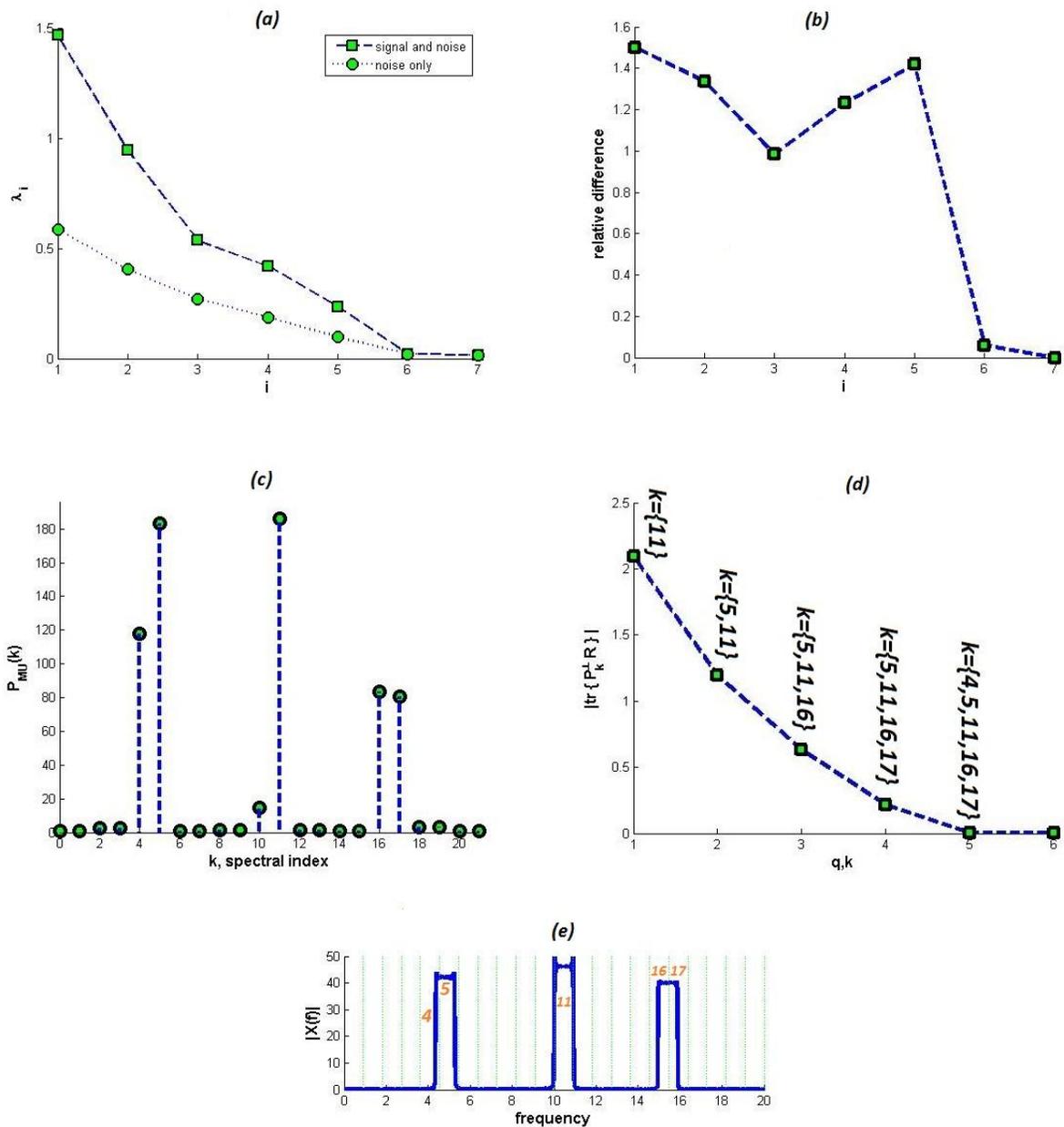

Fig. 3.13: (a) Ordered signal and theoretical noise eigenvalues, (b) relative error from the EFT algorithm (c) spectral index set by the MUSIC algorithm (d) spectral index set by the NLLS algorithm (e) frequency representation of the signal and its active cells

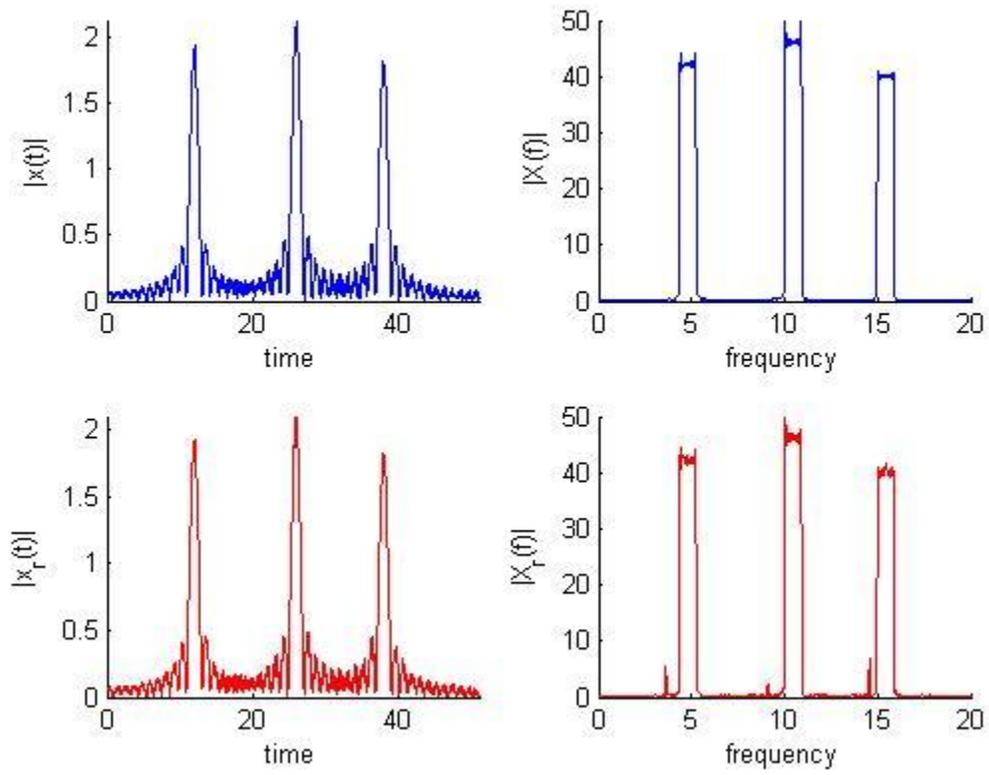

Fig.3.14: Time domain and frequency domain views of spectrum blind signal reconstruction

# 4. Application to cognitive radio

## 4.1 Introduction

Cognitive radio is a new paradigm for designing wireless communications systems, which aims to enhance the utilization of the radio frequency (RF) spectrum. The motivation behind cognitive radio is the scarcity of the available frequency spectrum and the increasing demand, caused by the emerging wireless applications for mobile users [25]. Fig.4.1 illustrates a cognitive network that contains primary or licensed users and secondary or cognitive users. The primary or licensed users are the systems that have been already assigned to a frequency band, whereas the secondary users are the systems that use the licensed bands when it is idle.

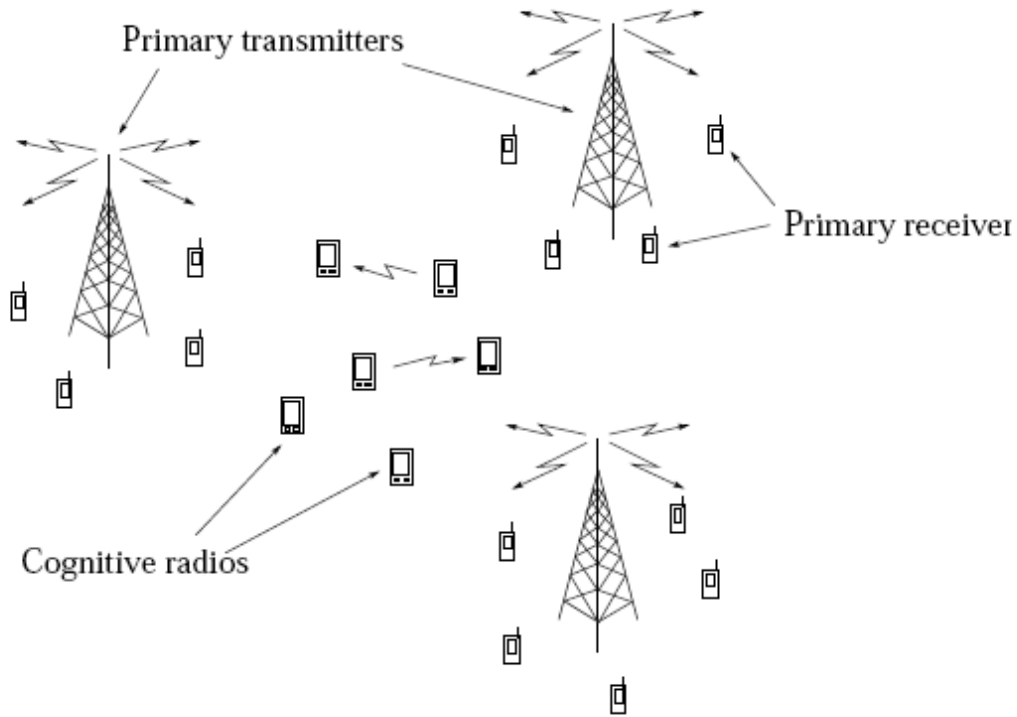

Fig.4.1: A network of cognitive radios that sense the radio frequency spectrum for spectrum opportunities and exploit them in an agile manner [31]

Due to the current static spectrum licensing scheme, *spectrum holes* or spectrum opportunities (Fig.4.2) arise. Spectrum holes are defined as frequency bands which are allocated to, but in some locations and at sometimes not utilized by, licensed users, and, therefore, could be accessed by unlicensed users [25].

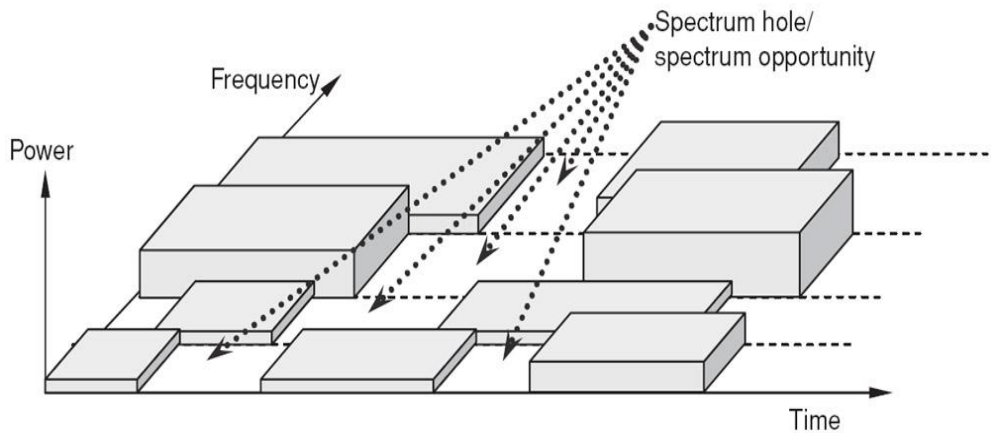

Fig.4.2: Spectrum holes or spectrum opportunity [25]

The main goal of cognitive radio is to provide adaptability to wireless transmission through dynamic spectrum access so that the performance of wireless transmission can be optimized, as well as enhancing the utilization of the frequency spectrum [25]. As such, the first cognitive task is to develop wireless spectral detection and estimation techniques for sensing and identification of the available spectrum [26].

## 4.2 Spectrum sensing

Spectrum sensing is an important function to enable CRs to detect the underutilized spectrum of primary systems and improve the overall spectrum efficiency [37]. Some well-known spectrum sensing techniques are energy detection, matched filter and cyclostationary feature detection that have been proposed for narrow band sensing. In all of these techniques the received signal is filtered with narrowband band-pass filters, sampled uniformly at Nyquist rate and then one of the above techniques is applied to decide between the two hypotheses $H_0$ and $H_1$. The hypothesis $H_0$ represents the case that no primary user is present and $H_1$ represents that a primary user exists. Fig. 4.3 shows the general implementation of narrow band spectrum sensing with conventional methods [37].

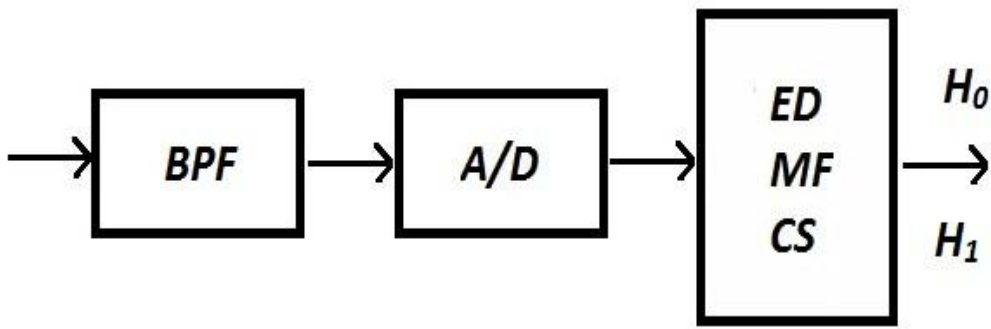

Fig.4.3: Conventional narrowband spectrum sensing techniques: Energy Detection, Matched Filter, Cyclostationary Detection

Future cognitive radios should be capable of scanning a wide band of frequencies, in the order of few GHz [27]. In the wideband regime, the radio front-end can employ a bank of band-pass filters to search a frequency band and then exploit the existing techniques for each narrow band, but this method requires a large number of RF components [26].

A conventional approach in wideband sensing is wavelet detection [25]. In order to identify the locations of vacant frequency bands, the entire wideband is then modeled as a train of consecutive frequency sub-bands where the power spectral characteristic is smooth within each sub-band, but changes abruptly on the border of two neighboring sub-bands. By employing a wavelet transform of the power spectral density (PSD) of the observed signal $x[n]$, the singularities of the PSD, $S(f)$, can be located and thus the vacant frequency bands can be found [37]. The digital implementation of a wavelet detector for spectrum sensing is shown in Fig.4.4. The major implementation challenge then lies in the very high sampling rates and high resolution *ADCs* with large dynamic range which have to operate at or above the Nyquist rate [26].

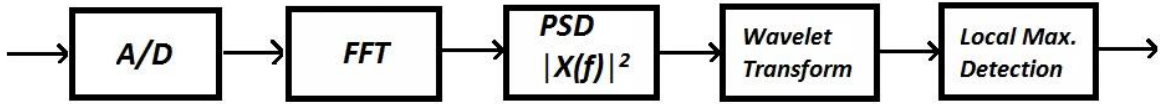

Fig.4.4: Digital implementation of a wavelet detector [37]

By using the fact that the wireless signals in open-spectrum networks are typically sparse in the frequency domain, the articles [26] and [27] introduced a compressive wide-band spectrum sensing with a combination of compressed sensing and wavelet transform. In this approach the received analog signal at the cognitive radio sensing receiver is transformed in to a digital signal using an analog-to-information (AIC) converter. The autocorrelation of this compressed signal is then used to reconstruct an estimate of the signal spectrum [27]. However as we mentioned above this approach is limited for detection of smooth spectral signals.

As considered in this project, the periodic non-uniform sampling can be used to overcome the problem of high sampling rate. Moreover, we showed how to estimate the spectral index set from the obtained compressed samples. In this way we introduce a wideband spectrum sensing model based on non-uniform sampling with sampling rates well below the Nyquist and effective for a wide range of signals [30].

The proposed architecture for spectrum sensing part of cognitive radio is illustrated in Fig.4.5. The operation of the system is as follows:

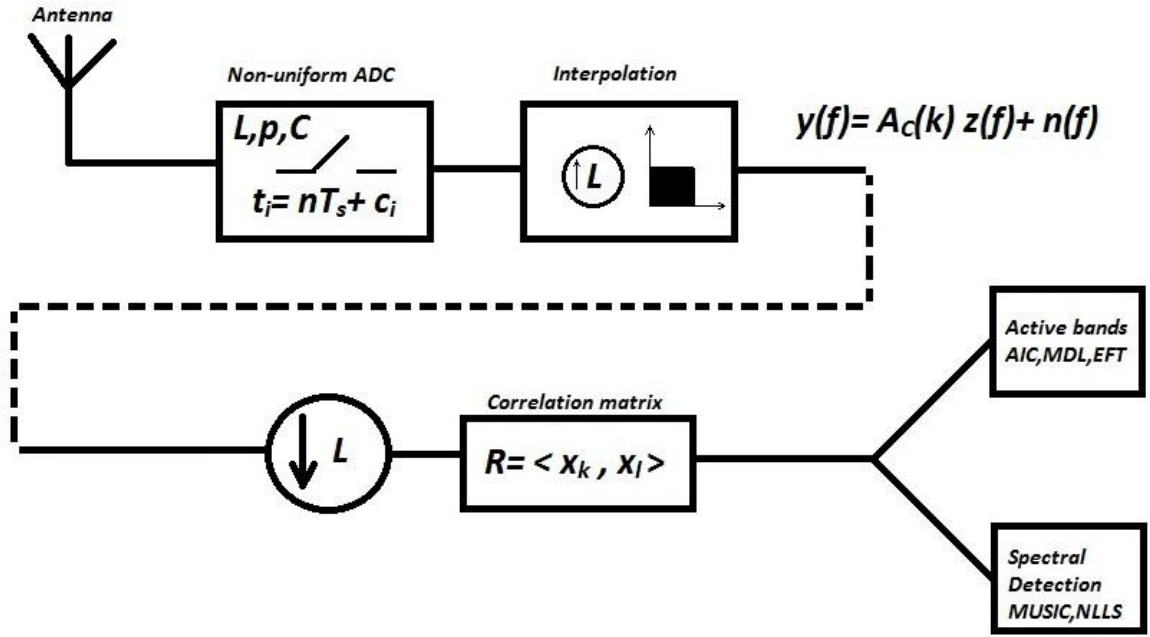

Fig.4.5: Proposed wideband spectrum sensing for cognitive radio based on Non-Uniform sampling

The analog received signal at the cognitive radio antenna is directly discredited by the Non-uniform ADC at a lower sample rate $f_s = (p/L) f_{max}$, compared with the maximum input frequency $f_{max}$. One possible implementation of the Non-uniform ADC is illustrated in Fig. 1.4. It is composed of $p$ parallel ADCs that each work uniformly at a sample rate of $f_s/p$ with different trigger times that are specified by the sample pattern $C$. In the articles [32],[33],[34] other architectures and practical issues are discussed. The Non-uniform ADC provides $p$ sequences of uniform sampled data where each sequence is configured according to (1.11) and then low-pass filtered with $f_{cut} = \frac{f_{max}}{L}$ at interpolation stage. The signal model at this point in the frequency domain was shown to be (3.8)

$$\mathbf{y}(f) = \mathbf{A}_C(\mathbf{k}) \, \mathbf{z}(f) + \mathbf{n}(f)$$

THis is a well-known classical signal model with some solutions based on the correlation matrix. For reducing the cost of computations, the sequences are down-sampled with a factor of $L$ and then using (3.20) the correlation matrix is computed. After computing the correlation matrix $\mathbf{R}$, as we discussed in the spectral recovery of blind spectrum signal, the number of active channels and also the location of them is determined by effective methods such as *MUSIC* or *NLLS*. Finally, the complement set of the recovered spectrum is the holes of spectrum and specifies the locations that the cognitive radio can start using for transmission.

**4.3 Specification and simulation of a cognitive radio network:**
Assume the cognitive network of Fig.4.1, with some primary users in the area and unknown frequency operation that can be placed anywhere inside the range of interest. The wideband system that should be the scanned by cognitive radio is assumed to be a frequency range of $[0, f_{max}]$ with occupancy of $\Omega$. Depending on the resolution $B$ (specified by application), the considered system is divided into $L = \frac{f_{max}}{B}$ spectral bands or channels. From (3.1) the number of active channels is in the interval

$$L\Omega \leq q \leq 2L\Omega \quad (4.1)$$

However the maximum number of $q$ is $2L\Omega$, but for large enough of $L$ the number of simultaneously active channels can be estimated by

$$\hat{q} \approx L\Omega \quad (4.2)$$

and then with choosing $p = \hat{q} + 1$ the system will work fine. This also suggests that the sampling rate is reduced by the factor of the system occupancy as

$$f_s = \frac{p}{L} f_{max} \approx \Omega f_{max} \quad (4.3)$$

Fig. 4.6 illustrates the spectrum of a typical cognitive system in the range of $[0, 2]GHz$ with three primary users respectively, a TV channel [470-500] *MHz*, GSM-900 uplink [824-849] *MHz* and downlink [869-894] *MHz* and a DECT system [1880-1900] *MHz*.

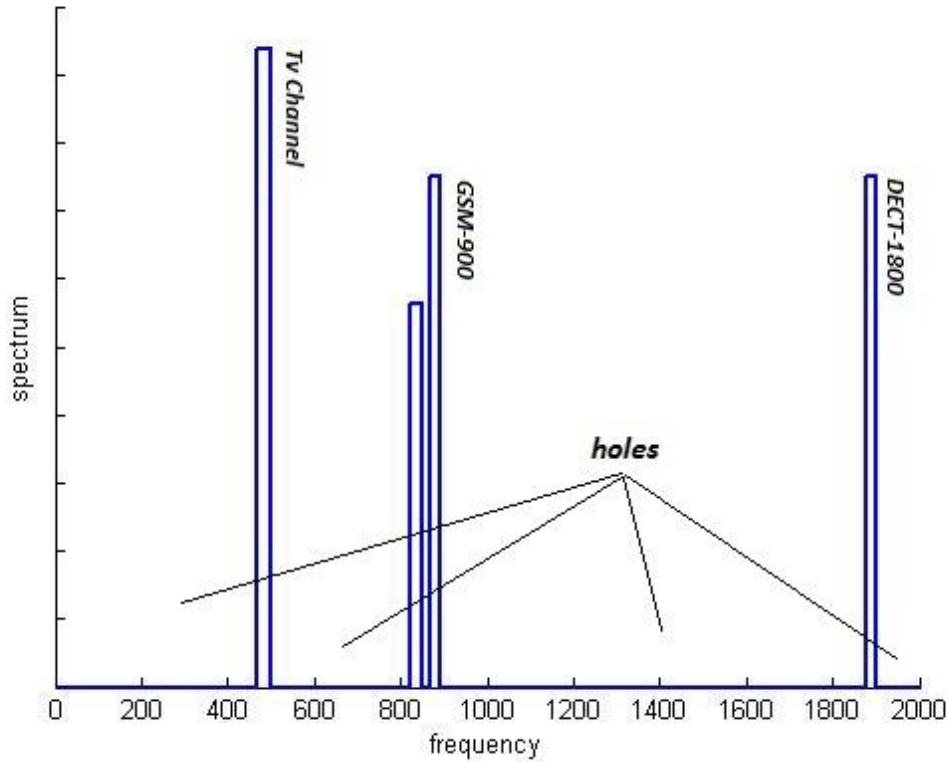

Fig. 4.6: Spectrum of a signal received at a cognitive radio, with three primary users occupying the corresponding bands

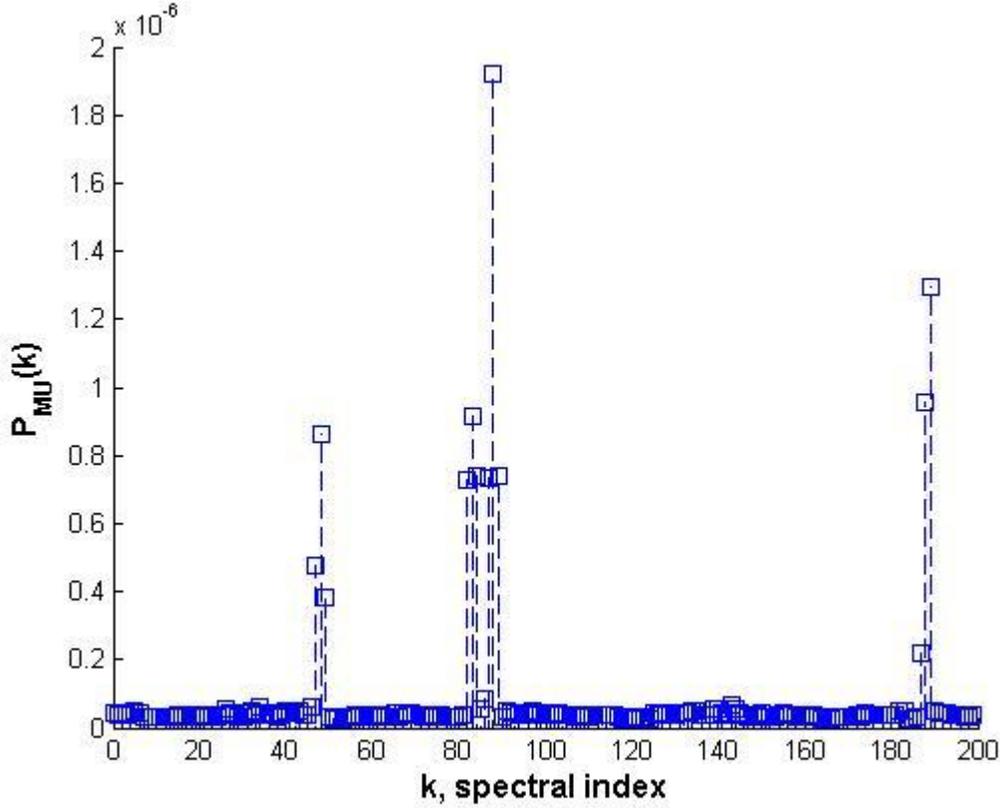

Fig.4.7: The occupied channels of the wideband system are detected by spectrum sensing

The cognitive radio needs to detect the holes of the spectrum with a resolution of $B=10MHz$, so $L=2GHz/10MHz=200$ should be chosen. The worst case occupancy is assumed to be $\Omega=0.1$, and then $p=200*0.1=20$ is selected for the Non-uniform ADC. The up-sampling is done with a factor of 200, and then the signal is filtered at $f_c=2GHz/200=10MHz$. Next, the down-sampling with a factor of 200 is applied. After computing the correlation matrix, the spectral recovery obtained with the MUSIC algorithm is shown in Fig.4.7. The indexes that do not have a peak can be treated as a free channel, and the location of the channel is in the interval

$$F_i = \left[\frac{i}{L}B : \frac{i+1}{L}B\right], \quad i \in \mathbf{k}_{free}, \quad \mathbf{k}_{free} = \mathbb{L} - \mathbf{k}$$

(4.4)

where the $\mathbf{k}_{free}$ is the complement of the spectral index set.

The performance of the method is evaluated by computing the probability of detection $P_d$, that is the probability of correct detection of an active channel:

$$P_d = P_r(\hat{k} = k)$$

(4.5)

The $P_d$, depends on the *SNR* of the channel and the compression ratio. The compression ratio (*CR*) is the factor of sampling frequency reduction as

$$CR = \frac{p}{L}$$
(4.6)

Hence, we set up a simulation test with a single frequency input as in Fig.4.8, with different *SNR* and compression ratio of *CR=0.1,0.2,0.3* and compute the $P_d$. The result of the simulation is shown in Fig.4.8. As the compression ratio increases, a higher $P_d$ is achieved at a lower *SNR*.

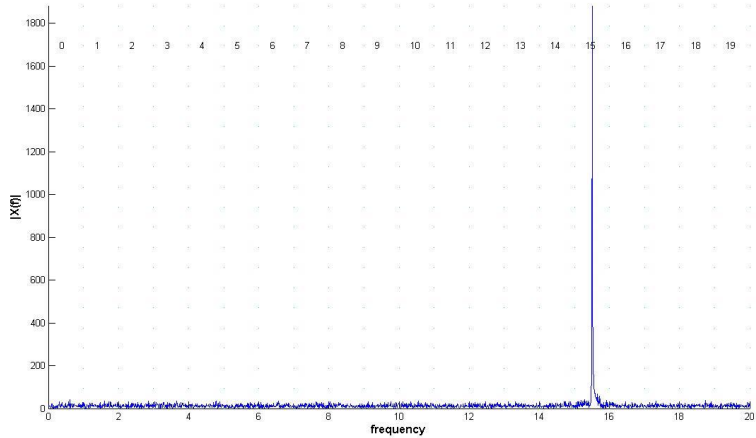

Fig.4.8: Frequency domain of input signal for probability detection test, *L=20*

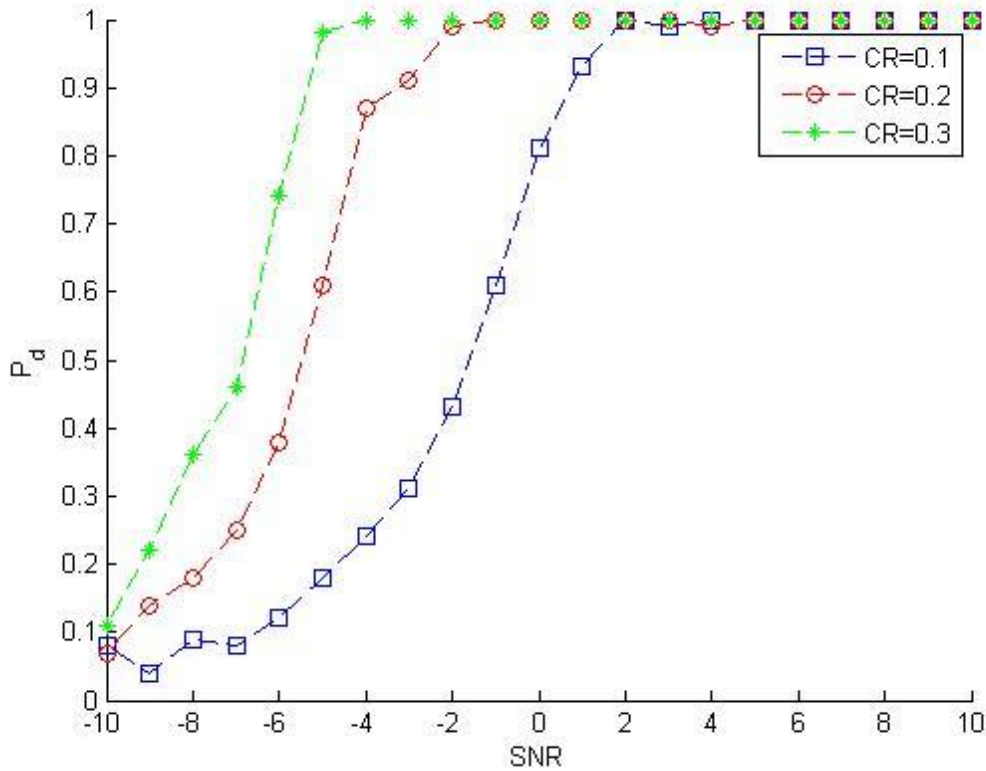

Fig.4.9: Probability of detection versus *SNR* and *CR*, $P_d$ increases with *CR*.

## 5- Summary and conclusions

This thesis investigates a clever way of sampling named multi-coset sampling to achieve a lower rate than the Nyquist rate for sampling of multiband signals. The Landau lower bound can be achieved by proper selection of sampling parameters. The model of the sampled data and the reconstruction formula were given. The relative reconstruction error for a typical case is found to be around 2.5%. The sampling parameters and their effect on the reconstructed signal were discussed. One of the most important parameters is the sample pattern. We propose an algorithm to find a suitable sample pattern. For unknown spectrum signals the methods of model order selection and spectral estimation based on the sampled data and the provided model were considered. The result of the model order process and spectral recovery depends on the number of sampled data, SNR and the compression ratio and is almost reliable even for low SNR, small number of sampled data or compression ratio.

Following the vision of sampling at lower rate and spectral recovery from fewer data, we noticed that the new paradigm of cognitive radio is looking for efficient and fast methods for wideband spectrum sensing. Hence we introduced a method for wideband spectrum sensing for cognitive radios based on periodic non-uniform sampling and evaluate it in terms of SNR and compression ratio.

Moreover, future studies can be focused on the issues such as: the implementation aspects of non-uniform analog to digital converters, formulation of a general sample pattern for any signal, consideration of general blind spectrum signal recovery with less information about the signal spectral support, and consideration of signals where the number of active slots is equal to the parameter $L$, but they have a low occupancy. Also, the detection probability can be theoretically investigated in terms of the SNR and compression ratio.